\documentclass[sigconf,nonacm,balance=false]{acmart}

\usepackage{mathtools}
\usepackage{multirow,array,subcaption}
\usepackage{xspace}
\usepackage{cleveref}
\usepackage{balance}
\ifxetex
  \AtBeginDvi{}
\fi

\setcopyright{none}
\acmDOI{}
\acmISBN{}

\definecolor{vargblue}{HTML}{165DFF}
\definecolor{vargteal}{HTML}{007F86}
\definecolor{varggreen}{HTML}{2A7F62}
\definecolor{vargred}{HTML}{B33A3A}
\definecolor{varggray}{HTML}{F3F6FA}
\newcommand{\method}{VARG\xspace}
\newcommand{\E}{\mathbb{E}}
\newcommand{\ind}{\mathbb{I}}
\newcommand{\gain}[1]{\textcolor{varggreen}{\textbf{#1}}}
\newcommand{\loss}[1]{\textcolor{vargred}{#1}}

\title[VARG: Value-Aware and Ranking-Aligned Generative Retrieval]{VARG: Value-Aware and Ranking-Aligned Generative Retrieval for Dynamic E-commerce Search}
\author{Xiaopeng Chu}
\affiliation{%
  \institution{University of Science and Technology of China}
  \city{Hefei}
  \state{Anhui}
  \country{China}}
\email{cxphhh@mail.ustc.edu.cn}

\author{Jianbo Zhu}
\affiliation{%
  \institution{Taobao \& Tmall Group of Alibaba}
  \city{Hangzhou}
  \country{China}}
\email{zhujianbo.zjb@taobao.com}

\author{Mingmin Jin}
\affiliation{%
  \institution{Taobao \& Tmall Group of Alibaba}
  \city{Hangzhou}
  \country{China}}
\email{jimmy.jmm@taobao.com}

\author{Jing Wang}
\affiliation{%
  \institution{Taobao \& Tmall Group of Alibaba}
  \city{Hangzhou}
  \country{China}}
\email{jing.wangj1@taobao.com}

\author{Xing Fang}
\authornote{Corresponding author.}
\affiliation{%
  \institution{Nankai University}
  \city{Tianjin}
  \country{China}}
\affiliation{%
  \institution{Taobao \& Tmall Group of Alibaba}
  \city{Hangzhou}
  \country{China}}
\email{fangxing.fx@taobao.com}

\author{Wenyi Zhang}
\affiliation{%
  \institution{University of Science and Technology of China}
  \city{Hefei}
  \state{Anhui}
  \country{China}}
\email{wenyizha@ustc.edu.cn}

\renewcommand{\shortauthors}{Chu et al.}

\begin{document}
\raggedbottom
\widowpenalty=10000
\displaywidowpenalty=10000
\begin{abstract}
Integrating recall and pre-ranking in e-commerce search requires candidate generation to account for relevance, personalization, and business value before final ranking. To this end, we present \method, a generative retrieval system for Tmall App search that directly admits generated item candidates to the existing final ranker. VARG-ID constructs semantic prefixes using RQ-VAE, enhances search relevance through bidirectional query--item contrastive learning, and combines these prefixes with a value-ordered third token to provide fine-grained item addresses and a business-value prior. Three-stage supervised fine-tuning progressively learns item-to-identifier mappings, query-semantic retrieval, and personalized retrieval. Personalized model training combines value-aware and hierarchy-aligned supervision with expanded user context, and uses local ordinal supervision (LO-SFT) to learn the local within-cluster ordering encoded by the third token. Prefix-GRPO combines gated rewards based on output legality, user behavior, ranker advantage, and search relevance with prefix-aware token weighting to align candidate generation with business value and ranking objectives. Coordinated daily product and model updates preserve existing item addresses while incorporating new products and behavioral feedback. Offline experiments on tens of millions of products validate identifier stability and demonstrate gains in retrieval quality and head-level value recall from SFT strategies and Prefix-GRPO over their respective baselines. In a 14-day online A/B test covering 20\% of search traffic, \method directly admits generated candidates to the final ranker and improves GMV by 1.45\%, per-user IPV by 0.22\%, and PCTR by 0.31\%. Online shopping-guide query evaluations further show that \method maintains competitive relevance with a smaller candidate quota.
\end{abstract}

\ccsdesc[500]{Information systems~Information retrieval}
\ccsdesc[300]{Information systems~Retrieval models and ranking}
\keywords{Generative Retrieval, E-commerce Search, Recall and Pre-ranking Integration, Semantic Identifiers, Reinforcement Learning}
\maketitle

\begin{figure}[t]
\centering
\captionsetup{skip=4pt}
\includegraphics[width=0.8\columnwidth]{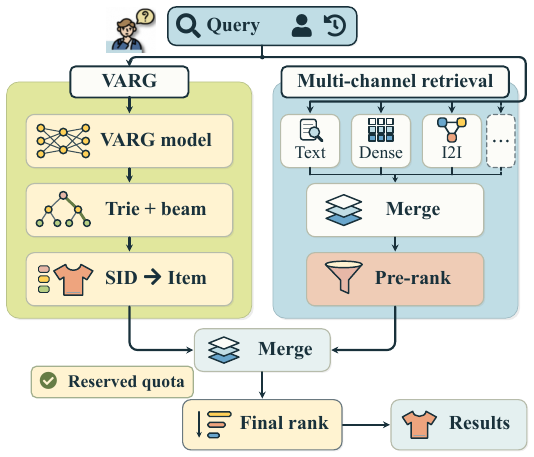}
\caption{Online integration of \method. Conventional multi-channel candidates undergo pre-ranking; VARG generates candidates using Trie-constrained beam search and SID-to-item mapping, bypassing pre-ranking via a reserved quota. Both streams are merged and deduplicated before final ranking.}
\Description{Online integration of \method. Conventional multi-channel candidates undergo pre-ranking; VARG generates candidates using Trie-constrained beam search and SID-to-item mapping, bypassing pre-ranking via a reserved quota. Both streams are merged and deduplicated before final ranking.}
\label{fig:onlinepath}
\end{figure}

\section{Introduction}
\label{sec:intro}

E-commerce search is a core information-access service in the Web ecosystem, and matches users' shopping needs to a massive set of products updated daily through a recall--pre-rank--rank cascade~\cite{zhu2026cq,chen2026onesearchv2}. Lexical retrieval~\cite{robertson2009bm25}, dense encoders~\cite{karpukhin2020dpr}, efficient nearest-neighbor indexes~\cite{malkov2016hnsw}, and late interaction~\cite{khattab2020colbert} offer complementary relevance and efficiency trade-offs. Recall and pre-ranking jointly perform candidate selection, accounting for both item relevance and business value. Since downstream rankers cannot recover items omitted from the candidate set, candidate selection should retain high-quality items under strict latency and resource budgets.

Generative retrieval formulates candidate selection as identifier generation directly from context~\cite{li2024generativesurvey}, with related work extending from document retrieval~\cite{tay2022dsi,wang2022nci} to hierarchical item IDs~\cite{rajput2023tiger}. Recent industrial systems further explore unified retrieval and ranking~\cite{deng2025onerec,zhou2025onerecv2} and intent-enhanced search~\cite{chen2026onesearchv2}. This formulation jointly models query semantics, user preferences, and behavioral context, expressing item selection and candidate priorities through generation probabilities.

Our setting is \emph{recall-and-pre-ranking integration}: an additional generative channel forwards items directly to the existing final ranker rather than replacing the search cascade (Figure~\ref{fig:onlinepath}). Bypassing item-level pre-ranking requires generated candidates to provide relevance, personalization, and business value within the reserved quota. The system also needs to accommodate ongoing daily product and model updates.

This goal raises three connected challenges.

\emph{Item-level selection and addressing stability.} Cluster-level identifiers improve item coverage during generation~\cite{zhu2026cq}, but shared identifiers limit the model's ability to distinguish individual items within a cluster. When item-level pre-ranking is bypassed, less relevant or lower-value items may occupy reserved candidate slots. Business-ranked identifiers provide finer addresses and value priors~\cite{ling2026crid}, but noisy conversion estimates and daily reordering can change existing item addresses and disrupt the model's learned addressing relationships. Learning new products also risks forgetting existing knowledge~\cite{mehta2022dsipp}.

\emph{Supervised learning for contextual retrieval.} The model needs to learn item-to-identifier mappings and select items according to query intent and user preferences. The challenge is to extend item memorization into query-semantic and personalized retrieval while ensuring that supervision reflects the semantic hierarchy and within-cluster ordering encoded by the identifiers~\cite{wang2024letter,chen2026onesearchv2}.

\emph{Business-value and ranking alignment.} An item's conversion propensity varies across queries and users, while learning only from observed target items provides limited guidance for comparing the quality and value of other candidates. Policy optimization offers a way to further adjust candidate priorities~\cite{deng2025onerec,zhu2026cq}, but sparse click and purchase matches poorly distinguish unobserved candidates. Training feedback should therefore also cover candidate relevance and downstream ranking performance while preventing invalid outputs from receiving positive credit.

\begin{figure*}[t]
\centering
\includegraphics[width=0.96\textwidth]{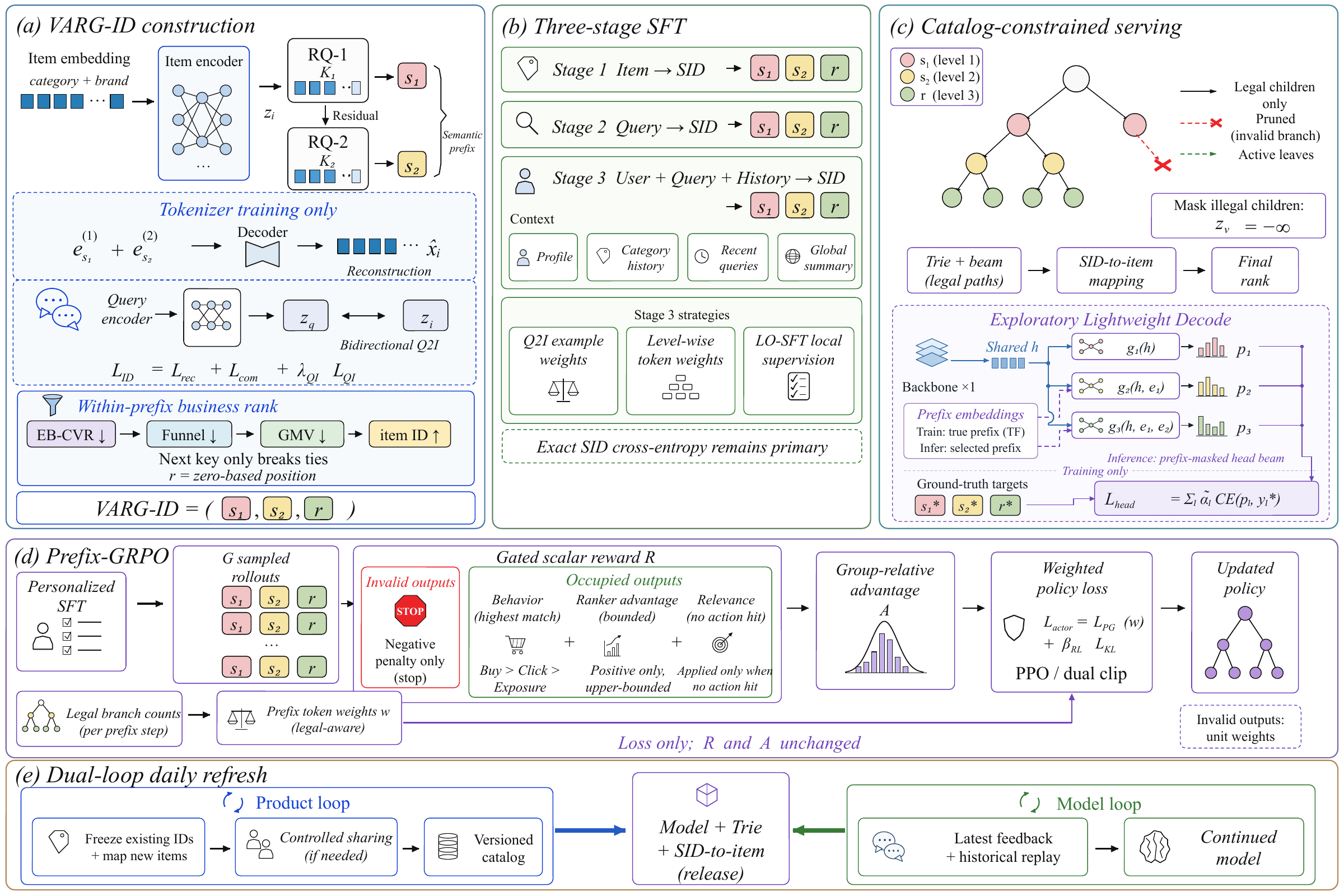}
\caption{Overview of \method. (a) VARG-ID with query-aligned semantic prefixes and value-ordered item addresses. (b) Three-stage SFT with context expansion and supervision strategies. (c) Trie-constrained serving and exploratory lightweight decoding. (d) Prefix-GRPO with gated multi-source rewards and prefix-aware token weighting. (e) Coordinated daily product and model updates.}
\Description{Overview of \method. (a) VARG-ID with query-aligned semantic prefixes and value-ordered item addresses. (b) Three-stage SFT with context expansion and supervision strategies. (c) Trie-constrained serving and exploratory lightweight decoding. (d) Prefix-GRPO with gated multi-source rewards and prefix-aware token weighting. (e) Coordinated daily product and model updates.}
\label{fig:framework}
\end{figure*}

To address these challenges, we present \method (Figure~\ref{fig:framework}) with the following main contributions:
\begin{itemize}
  \item We introduce \emph{VARG-ID}, a hierarchical item identifier with collision-free initial construction. Bidirectional Q2I contrastive learning injects search relevance into its semantic prefix, while stable empirical-Bayes conversion-rate ordering encodes business value in its third token. We further establish stable daily-refresh mechanisms for both items and the retrieval model.
  \item We build a three-stage SFT pipeline that progresses from item memorization to query-semantic retrieval and personalized retrieval. Personalized training combines Q2I value weighting, SID level-wise loss, and expanded user context, and uses local ordinal supervision (LO-SFT) to learn the within-cluster ordering encoded by the third token. We also explore lightweight parallel decoding to reduce inference cost.
  \item We develop Prefix-GRPO with gated multi-source rewards over SID legality, user behavior, ranker advantage, and search relevance, providing denser reward signals for policy optimization. Combined with prefix-aware token weighting, these rewards align candidate priorities with business value and downstream ranking objectives.
\end{itemize}

Offline experiments evaluate identifier stability and the effects of SFT and Prefix-GRPO on retrieval quality and value recall. A 14-day online A/B test validates the deployed recall-and-pre-ranking system, yielding a 1.45\% GMV gain.

\section{Related Work}
\label{sec:related}

\subsection{Generative Retrieval and Training}
Generative retrieval uses different forms of identifiers: GENRE generates entity names~\cite{decao2021genre}, DSI and NCI generate document identifiers~\cite{tay2022dsi,wang2022nci}, and SEAL generates indexed document substrings~\cite{bevilacqua2022seal}. Wu et al.~\cite{wu2024multivector} connect generative retrieval to multi-vector dense retrieval through a shared relevance-scoring framework. TIGER~\cite{rajput2023tiger} extends this paradigm to recommendation with hierarchical semantic IDs.

For generative retrieval training, P5~\cite{geng2022p5} unifies recommendation tasks through text-to-text learning and personalized prompts, while LC-Rec~\cite{zheng2024lcrec} aligns language and collaborative semantics around quantized item indices. Industrial systems further explore unified retrieval and ranking~\cite{deng2025onerec,zhou2025onerecv2}: CQ-SID~\cite{zhu2026cq} progressively learns item, query, and personalized mappings, OneSearch-V2~\cite{chen2026onesearchv2} models query intent through keyword-based reasoning and self-distillation, and TBGRecall~\cite{liang2025tbgrecall} combines multi-session modeling with incremental training for e-commerce recommendation. These studies provide a foundation for item representation, staged training, and context modeling. For integrated recall and pre-ranking, a further consideration is whether the generative model can select items that balance relevance, personalization, and business value within a limited candidate quota.

\subsection{Semantic IDs and Dynamic Indexing}
Vector and residual quantization provide a discrete representation basis for hierarchical identifiers~\cite{oord2017vqvae,lee2022rqvae,rajput2023tiger}. Subsequent studies further integrate identifier learning with retrieval objectives: GenRet~\cite{sun2023genret} jointly learns document tokenization, reconstruction, and retrieval; IDGenRec~\cite{tan2024idgenrec} learns textual IDs with a recommender; and LETTER~\cite{wang2024letter} combines semantic, collaborative, and diversity regularization with ranking-guided generation. D$^2$-DocID~\cite{cheng2025d2docid} uses contrastive learning to build descriptive identifiers that distinguish similar documents, while FORGE~\cite{fu2026forge} studies industrial SID design using intrinsic evaluation measures.

In search, CQ-SID~\cite{zhu2026cq} injects query relevance while retaining cluster-level aggregation, and CRID~\cite{ling2026crid} uses intra-cluster business ranking for item-level addressing and incremental updates. For dynamic corpora, DSI++~\cite{mehta2022dsipp} uses generative replay to mitigate forgetting, while CLEVER~\cite{chen2023clever} combines incremental product quantization with memory-augmented learning. These studies involve two forms of stability: preserving model knowledge and maintaining stable item identifiers. Under daily product updates, identifier design should balance fine-grained addressing, business value, and decoding efficiency while reducing the disruption to existing addressing relationships caused by changes in item statistics.

\subsection{Policy Optimization for Retrieval}
Ranking-oriented training further considers the quality and ordering of generated candidates. LTRGR~\cite{li2024ltrgr} adds a ranking-oriented training phase without extra inference stages, while relevance-feedback RL~\cite{zhou2023relevancefeedback} aligns token-level generation with document-level relevance. PPO, DPO, and GRPO provide general frameworks for policy or preference optimization~\cite{schulman2017ppo,rafailov2023dpo,shao2024deepseekmath}.

In recommendation, OneRec~\cite{deng2025onerec} uses reward-model-derived preference pairs for iterative DPO, and OneRec-V2~\cite{zhou2025onerecv2} incorporates real interaction feedback. In search, EG-GRPO~\cite{zhu2026cq} injects expert SIDs to alleviate sparse positive feedback, while OneSearch-V2~\cite{chen2026onesearchv2} combines behavioral and relevance rewards with token-position marginal advantages. These methods demonstrate the value of additional feedback for improving generative retrieval. For hierarchical SID generation, we further study how to differentiate legal candidates without behavioral feedback, prevent invalid outputs from receiving positive credit, and distribute supervision across decisions at different levels to improve ranking quality within a limited candidate quota.

\section{Method}
\label{sec:method}

\paragraph{Problem setting.}
Given a query $q$, user profile $u$, and history $h$, a Transformer~\cite{vaswani2017attention} conditioned on context $x=(q,u,h)$ autoregressively generates a three-token item identifier $y=(s_1,s_2,r)$. At inference, Trie-constrained beam search produces a ranked candidate list $\mathcal{Y}_K$ of $K$ SIDs, which are expanded into item candidates through the SID-to-item map and forwarded directly to the existing final ranker as an additional retrieval channel. Under a fixed decoding budget and final-ranking candidate quota, the model should generate valid SIDs and select items that are relevant, personalized, and commercially valuable. The item map and model should also accommodate daily product updates and new user feedback.

\subsection{VARG-ID: Semantics Followed by Stable Value}
\label{sec:vargid}

A VARG-ID consists of a semantic prefix and a value-ordered token: the query-aligned prefix supports semantic routing, while the third token provides fine-grained within-prefix item addresses and a business-value prior.

\paragraph{Semantic-prefix construction.}
The item embedding $x_i\in\mathbb{R}^{d_x}$ incorporates category and brand attributes and is mapped to $z_i\in\mathbb{R}^{d_z}$ by the item encoder $E_I$. The query encoder $E_Q$, used only during tokenizer training, maps the paired search query to $z_q\in\mathbb{R}^{d_z}$. Two codebooks of sizes $K_1$ and $K_2$ perform residual quantization sequentially:
\begin{align}
 s_{1,i}&=\arg\min_k\|z_i-e^{(1)}_k\|_2^2,
 &\delta_i^{(1)}&=z_i-\operatorname{sg}(e^{(1)}_{s_{1,i}}),\notag\\
 s_{2,i}&=\arg\min_k\|\delta_i^{(1)}-e^{(2)}_k\|_2^2,
 &z_i^{q}&=e^{(1)}_{s_{1,i}}+e^{(2)}_{s_{2,i}}.
 \label{eq:rq}
\end{align}
Here $\operatorname{sg}$ denotes stop-gradient. The decoder uses the straight-through estimate $z_i^{\mathrm{ST}}=z_i+\operatorname{sg}(z_i^q-z_i)$, whose forward value equals $z_i^q$, while reconstruction gradients flow to the item encoder.

\begingroup
\predisplaypenalty=0
\paragraph{Query-aligned tokenizer objective.}
The tokenizer jointly optimizes reconstruction, commitment, and bidirectional query--item contrastive losses:
\begin{align}
 \mathcal{L}_{\mathrm{rec}}
 &=\|D(z_i^{\mathrm{ST}})-x_i\|_2^2,\notag\\
 \mathcal{L}_{\mathrm{com}}
 &=\tfrac12\!\left(\|z_i-\operatorname{sg}(e^{(1)}_{s_{1,i}})\|_2^2
 +\|\delta_i^{(1)}-\operatorname{sg}(e^{(2)}_{s_{2,i}})\|_2^2\right),\notag\\
 \mathcal{L}_{\mathrm{QI}}
 &=\tfrac12\left[\operatorname{CE}(S,I)+\operatorname{CE}(S^\top,I)\right],\quad
 S_{ab}=\frac{\bar z_{i,a}^{\top}\bar z_{q,b}}{\tau_{\mathrm{QI}}},\notag\\
 \mathcal{L}_{\mathrm{ID}}
 &=\mathcal{L}_{\mathrm{rec}}+\mathcal{L}_{\mathrm{com}}
 +\lambda_{\mathrm{QI}}\mathcal{L}_{\mathrm{QI}}.
 \label{eq:idloss}
\end{align}
\endgroup
Reconstruction preserves item embedding information, commitment encourages encoder outputs to approach discrete codes, and the bidirectional InfoNCE term brings paired query and item representations closer in continuous space, injecting search relevance before quantization. Here, $\tau_{\mathrm{QI}}>0$ is the contrastive temperature and $\lambda_{\mathrm{QI}}>0$ is the contrastive-loss weight. The diagonal of the contrastive matrix $S$ contains positive pairs, with labels $I=(0,\ldots,N-1)$; the remaining valid query--item combinations in the local minibatch serve as negatives. Samples without queries optimize only the reconstruction and commitment losses. The codebooks are initialized by deterministic K-means and updated by exponential moving averages.

The first two codes form the semantic prefix $c(i)=(s_{1,i},s_{2,i})$. At initial construction, the third token $r_i$ is the item's value rank within that prefix, yielding a unique item address:
\begin{equation}
  y_i=(s_{1,i},s_{2,i},r_i).
  \label{eq:vargid}
\end{equation}
Here $r_i$ is zero-based and satisfies $0\leq r_i<K_3$, where $K_3$ is the rank-token vocabulary size.

\paragraph{Value-ordered item address.}
Let $b_i$, $a_i$, and $c_i$ denote an item's purchase, add-to-cart, and click counts in a fixed ranking window. We set $v_i=b_i+\eta_a a_i$ and estimate conversion value with empirical-Bayes smoothing:
\begin{equation}
 p_0=\frac{\sum_i v_i}{\sum_i c_i},\qquad
 \widehat{p}_i=\frac{v_i+\alpha_{\mathrm{ID}}p_0}{c_i+\alpha_{\mathrm{ID}}}.
 \label{eq:ebcvr}
\end{equation}
We define the behavioral score $V_i^{\mathrm{funnel}}=\gamma_b b_i+\gamma_a a_i+\gamma_c c_i$ and rank items within each prefix using the following funnel rule, comparing its keys lexicographically:
\begin{equation}
 \widehat p_i\downarrow
 \;\longrightarrow\;V_i^{\mathrm{funnel}}\downarrow
 \;\longrightarrow\;\mathrm{GMV}_i\downarrow
 \;\longrightarrow\;\mathrm{ID}_i\uparrow.
 \label{eq:rankfunnel}
\end{equation}
Here $0<\eta_a<1$, $\alpha_{\mathrm{ID}}>0$, and $\gamma_b>\gamma_a>\gamma_c>0$, reflecting the behavioral priority of purchase $>$ add-to-cart $>$ click. Empirical-Bayes smoothing reduces low-frequency noise while preserving conversion-value information; \Cref{app:sid-diagnostics} shows that its ordering is more stable than raw click and volume ordering.

\subsection{Dual-Loop Daily Refresh}
\label{sec:refresh}

\paragraph{Product loop.}
We rebuild SIDs monthly to create a new primary version. Between rebuilds, we freeze the RQ-VAE and SID vocabulary, admit new items daily, and keep assigned item SIDs fixed. New items in occupied prefixes are jointly ranked with existing items using current value statistics; their ranks select historical slots, capped at the last slot. New prefixes receive slots starting from zero. Once assigned, new item addresses also remain fixed within the version.

For example, existing items A, B, and C under one semantic prefix have third tokens 0, 1, and 2. If a new item D ranks between A and B by value, D reuses slot 1 and shares B's SID, while the addresses of A, B, and C remain unchanged.

For prefix $c=(a,b)$, $C_c^{(t)}=C_{ab}^{(t)}$ denotes the number of distinct occupied rank slots at snapshot $t$. An occupied prefix has maximum rank $M_t(c)=C_c^{(t)}-1$; unseen prefixes have capacity zero. Let $\rho_t(i)$ be the zero-based position of new item $i$ after jointly ordering all existing and new items in the same prefix at snapshot $t$ according to \Cref{eq:rankfunnel}. Its assigned third-token address is
\begin{equation}
 r_t(i)=
 \begin{cases}
   \min\{\rho_t(i),M_{t-1}(c(i))\}, & C_{c(i)}^{(t-1)}>0,\\
   \min\{\rho_t(i),K_3-1\}, & C_{c(i)}^{(t-1)}=0.
 \end{cases}
 \label{eq:incremental}
\end{equation}
For new prefixes, \Cref{eq:incremental} caps slot indices at $K_3-1$. Shared slots are expanded into item candidates for downstream ranking at serving time. Below, $C_{ab}$ denotes the slot capacity of the relevant fixed snapshot, with the time index omitted.

\paragraph{Model loop.}
Each day, the latest click and purchase data are labeled using the SID mapping in effect when the behavior occurred. Within the same primary SID version, we continue training the last accepted personalized generation model for $E_{\mathrm{day}}\in\mathbb{N}^{+}$ epochs, mixing historical replay with proportion $0\leq\lambda_{\mathrm{replay}}\leq1$ to incorporate fresh feedback while retaining learned addressing relationships. After validation, the model, Trie, and item map are atomically published as a compatible version combination.

Monthly SID rebuilding introduces a new primary version, under which training labels, SID-encoded histories, and replay data are re-encoded. The retrieval model is retrained before joint deployment, and the accepted model initializes the next daily-update cycle.

\subsection{Three-Stage SFT}
\label{sec:sft}


\paragraph{Three-stage supervised fine-tuning.}
We add $K_\ell$ dedicated tokens for SID level $\ell$ to the vocabulary of Qwen2.5-0.5B-Instruct~\cite{qwen2025}. Given context $x$, the model autoregressively generates the target identifier $y=(s_1,s_2,r)$:
\begin{equation}
 p_\theta(y\mid x)=\prod_{\ell=1}^{3}
 p_\theta(y_\ell\mid x,y_{<\ell}).
 \label{eq:arfactor}
\end{equation}
Training proceeds sequentially across stages. Stage~1 learns Item$\to$\allowbreak SID from product content, establishing item-address memory. Stage~2 continues from Stage~1 using Query$\to$SID examples to learn query-semantic retrieval. Stage~3 continues from Stage~2 using User+\allowbreak Query+\allowbreak History$\to$\allowbreak SID examples to learn personalized retrieval.

SFT Base is the personalized retrieval model after three-stage training, using the current query, user profile, and related-category click history with standard cross-entropy supervision. We separately compare context expansion, joint Q2I and level weighting, level weighting alone, and LO-SFT against this baseline. Prefix-GRPO (\Cref{tab:rl}) initializes its trainable policy $\pi_\theta$ and fixed reference policy $\pi_{\mathrm{ref}}$ identically from the SFT Base checkpoint (\Cref{tab:sft}).

\paragraph{User context expansion.}
The personalized input is $x=(q,\allowbreak u,\allowbreak h^{\mathrm{cat}},\allowbreak h^{\mathrm{recent}},\allowbreak g)$, where $q$ is the current query, $u$ is the user profile including gender and purchasing power, $h^{\mathrm{cat}}$ is related-category click history represented by historical SIDs, $h^{\mathrm{recent}}$ contains the $L_{\mathrm{recent}}$ most recent search queries, and $g$ is a global-interest summary distilled from cross-domain behaviors. Recent queries provide additional signals of short-term user intent, while the global summary represents long-term user interests by summarizing category, brand, and price preferences. Figure~\ref{fig:promptexample} shows an example of item candidate generation from the full input context.

\paragraph{Q2I example weighting.}
To account for conversion differences across query--item pairs, we construct example weights using purchase and click statistics from a fixed historical window. Let $b_j$ and $c_j$ denote the purchase and click counts for query--item pair $j$. The global prior, smoothed conversion rate, and example weights are
\begin{align}
 p_{0q}&=\frac{\sum_j b_j}{\sum_j c_j},
 &\widehat p_j&=\frac{b_j+\alpha_q p_{0q}}{c_j+\alpha_q},\notag\\
 \widetilde w_j&=\operatorname{clip}
 \left(\frac{\widehat p_j}{p_{0q}},w_{\min},w_{\max}\right),
 &w_j&=\frac{\widetilde w_j}{\E[\widetilde w]}.
 \label{eq:q2iw}
\end{align}
Here, $\alpha_q>0$ is the prior strength, and the clipping bounds satisfy $0<w_{\min}<1<w_{\max}$. Empirical-Bayes smoothing reduces noise in sparse statistics; the ratio of the smoothed conversion rate to the global prior is clipped and normalized by its mean, assigning greater weights to query--item pairs with higher conversion propensity. We activate this signal only in Stage~3; Stages~1--2 remain equally weighted to learn unbiased item memory and query semantics.

\paragraph{Level-aware token weighting.}
The three target positions have distinct decision roles. $s_1$ selects a coarse semantic region, $s_2$ chooses the fine semantic prefix, and $r$ only identifies an item after that prefix is correct. In Trie-constrained Top-1000 evaluation, the HR gap is 16.94 percentage points between L1 and L1+L2, versus only 3.80 points between L1+L2 and the complete SID, indicating that the second level is the main prediction bottleneck. We therefore assign supervision weights based on prediction performance at each level. Let $\ell_{j,t}=-\log p_\theta(y_{j,t}\mid x_j,y_{j,<t})$ denote per-token cross-entropy, $m_{j,t}$ the response mask, and $\lambda(y_{j,t})$ the token level. The level-weighted loss is
\begin{equation}
 \overline{\mathcal L}^{\mathrm{level}}_j=
 \frac{\sum_t m_{j,t}\alpha_{\lambda(y_{j,t})}\ell_{j,t}}
 {\sum_t m_{j,t}\alpha_{\lambda(y_{j,t})}}.
 \label{eq:levelloss}
\end{equation}
The level coefficients $(\alpha_1,\alpha_2,\alpha_3)$ satisfy $\alpha_2>\alpha_1>\alpha_3>0$ to strengthen supervision at the first two levels; non-SID response tokens retain unit weight. Combining this loss with Q2I example weights gives the Stage~3 training objective
\begin{equation}
 \mathcal L_{\mathrm{SFT3}}=
 \frac{\sum_j w_j\overline{\mathcal L}^{\mathrm{level}}_j}
 {\sum_j w_j}.
 \label{eq:sft3}
\end{equation}

\paragraph{Local ordinal supervision (LO-SFT).}
Standard cross-entropy treats rank tokens as discrete categories without explicitly using their ordinal distances. LO-SFT continues fine-tuning from the trained Stage-3 SFT Base checkpoint, adding local supervision around the target position based on the value ranks encoded by the third token.

For prefixes with slot capacity $C=C_{ab}\geq C_{\min}$, we construct a local target distribution centered on the true rank $r^*$ that decays with distance, while retaining the exact-token loss:
\begin{align}
 q(r\mid r^*)&=\frac{\exp(-|r-r^*|/\tau_{\mathrm{LO}})}{Z(r^*)}
 \ind[|r-r^*|\leq R_{\mathrm{LO}}]\notag\\
 &\quad\times\ind[0\leq r<C],\notag\\
 \mathcal{L}_{\mathrm{LO}}&=\mathcal{L}_{\mathrm{exact}}
 -\lambda_{\mathrm{LO}}\sum_{r=0}^{C-1}q(r\mid r^*)
 \log p_\theta(r\mid x,s_1,s_2),
 \label{eq:losft}
\end{align}
where $C_{\min}$ is the minimum prefix-capacity threshold, $R_{\mathrm{LO}}$ is the neighborhood radius, $\tau_{\mathrm{LO}}>0$ controls distance decay, $\lambda_{\mathrm{LO}}>0$ is the auxiliary-loss weight, and $Z(r^*)$ is the normalization constant over the legal neighborhood. The exact loss supervises the target SID, while the auxiliary term guides the model to learn the local ordinal relationships encoded by EB-CVR ordering.

\subsection{Prefix-GRPO}
\label{sec:grpo}

Building on supervised learning, Prefix-GRPO adjusts candidate generation probabilities according to rewards to further align candidate priorities with business value and ranking objectives. We use group-relative policy optimization~\cite{shao2024deepseekmath}, combining gated multi-source rewards with prefix-aware token weighting. For each prompt, the policy samples $G$ responses, with $(T_{\mathrm{roll}},p_{\mathrm{roll}},k_{\mathrm{roll}})$ controlling sampling temperature, nucleus sampling, and top-$k$ truncation, respectively.

\paragraph{Response contract and gated reward.}
After removing control tokens, a response must be exactly
$y=\langle s1_a\rangle\langle s2_b\rangle\langle s3_r\rangle$, where $0\leq a<K_1$, $0\leq b<K_2$, and $0\leq r<K_3$. In the RL training snapshot, a SID with $C_{ab}>0$ and $0\leq r<C_{ab}$ corresponds to an occupied slot. Only well-formed, occupied SIDs participate in subsequent reward computation; other outputs receive only their corresponding negative rewards.

Let $\mathcal P_x$, $\mathcal C_x$, and $\mathcal E_x$ denote the sets of SIDs matched to purchase, click, and exposure behaviors for request $x$, respectively. The behavior reward uses the highest matched action under the priority purchase $>$ click $>$ exposure:
\begin{equation}
 R_{\mathrm{UA}}(x,y)=
 \begin{cases}
 \beta_{\mathrm{buy}}, & y\in\mathcal P_x,\\
 \beta_{\mathrm{clk}}, & y\notin\mathcal P_x,\ y\in\mathcal C_x,\\
 \beta_{\mathrm{exp}}, & y\notin\mathcal P_x\cup\mathcal C_x,\ y\in\mathcal E_x,\\
 0, & \text{otherwise}.
 \end{cases}
 \label{eq:uareward}
\end{equation}
The ranker-advantage reward favors items with scores above the candidate mean for the same request. Let $f_x(y)$ be the production ranker's final score and $\mu_x^f,\sigma_x^f$ its mean and standard deviation over mapped recall candidates for that request. The reward is
\begin{align}
 a_x(y)&=\frac{f_x(y)-\mu_x^f}{\sigma_x^f},\notag\\
 \widetilde a_x(y)&=
 \begin{cases}
 a_x(y), & 0<a_x(y)<\infty,\\
 0, & \text{otherwise},
 \end{cases}\notag\\
 R_{\mathrm{adv}}(x,y)&=\beta_{\mathrm{adv}}\tanh
 \left(\frac{\widetilde a_x(y)}{\tau_{\mathrm{adv}}}\right).
 \label{eq:advreward}
\end{align}
For occupied SIDs without matched user behavior, the relevance grade $g_x(y)$ provides supplementary feedback:
\begin{equation}
 R_{\mathrm{rel}}(x,y)=
 \begin{cases}
 \beta_{\mathrm{rel}}^{(3)}, & g_x(y)=3,\\
 \beta_{\mathrm{rel}}^{(2)}, & g_x(y)=2,\\
 -\beta_{\mathrm{rel}}^{(1)}, & g_x(y)=1,\\
 0, & \text{otherwise}.
 \end{cases}
 \label{eq:relreward}
\end{equation}
Let $\mathcal F$ denote malformed or out-of-range outputs, $\mathcal U$ structurally valid but unoccupied outputs, and $\mathcal H_x=\mathcal P_x\cup\mathcal C_x\cup\mathcal E_x$. The complete reward is
\begin{equation}
 R(x,y)=
 \begin{cases}
 -\beta_{\mathrm{F}}, & y\in\mathcal F,\\
 -\beta_{\mathrm{U}}, & y\in\mathcal U,\\
 R_{\mathrm{UA}}+R_{\mathrm{adv}}, & y\text{ occupied},\ y\in\mathcal H_x,\\
 R_{\mathrm{adv}}+R_{\mathrm{rel}}, & y\text{ occupied},\ y\notin\mathcal H_x.
 \end{cases}
 \label{eq:reward}
\end{equation}
Here, $\beta_{\mathrm{F}}>\beta_{\mathrm{U}}>0$, $\beta_{\mathrm{buy}}>\beta_{\mathrm{clk}}>\beta_{\mathrm{exp}}>0$, $\beta_{\mathrm{adv}},\tau_{\mathrm{adv}}>0$, $\beta_{\mathrm{rel}}^{(3)}>\beta_{\mathrm{rel}}^{(2)}>0$, and $\beta_{\mathrm{rel}}^{(1)}>0$. Relevance rewards apply only to SIDs without matched behavior, avoiding double counting of correlated labels.

\paragraph{Prefix-aware token weighting.}
The number of legal token alternatives varies across prefixes. We use these counts to assign token-level policy-loss weights, giving higher weights to decisions with more branches. In the RL training snapshot, the numbers of legal child tokens at the second and third levels are
\begin{equation}
 n_2(a)=\sum_{b=0}^{K_2-1}\ind[C_{ab}>0],\qquad
 n_3(a,b)=C_{ab},
 \label{eq:branchcounts}
\end{equation}
Let $N_2=\max_a n_2(a)$ and $N_3=\max_{a,b}n_3(a,b)$. We map the legal child-token count to $[d,1]$ using
\begin{equation}
 \phi(n;N)=
 \begin{cases}
 d+(1-d)\dfrac{\log n}{\log N}, & N>1,\\
 d, & N=1,
 \end{cases}
 \qquad 0<d\leq1.
 \label{eq:prefixmap}
\end{equation}
For an occupied path, token weights are
\begin{align}
 w_{s_1}&=1, &w_{s_2}&=\phi(n_2(a);N_2),\notag\\
 w_{s_3}&=\phi(n_3(a,b);N_3), &w_{\mathrm{ctrl}}&=d.
 \label{eq:prefixweight}
\end{align}
The first-level weight is fixed at 1, while later levels are weighted according to the number of legal alternatives under the current prefix. Branches with only one legal option, as well as EOS and control tokens, use the floor $d$. For malformed or unoccupied responses, all valid tokens use unit weight. These weights enter the policy loss in \Cref{eq:weightedpg}.

\paragraph{Weighted GRPO optimization.}
Let $R_{x,g}$ be the reward of response $g$ to prompt $x$, and $\bar R_x,s_x$ the mean and sample standard deviation of the $G$ response rewards for that prompt. Based on GRPO's within-group reward comparison~\cite{shao2024deepseekmath}, we construct advantages through standardization, whitening, and clipping:
\begin{align}
 \widehat R_{x,g}&=\operatorname{clip}
 \left(\frac{R_{x,g}-\bar R_x}{s_x+\varepsilon_{\mathrm{num}}},-c_R,c_R\right),\notag\\
 A_{x,g}&=\operatorname{clip}
 \left(\operatorname{Whiten}(\widehat R_{x,g}),-c_A,c_A\right).
 \label{eq:grpoadv}
\end{align}
Here, $\varepsilon_{\mathrm{num}}>0$ provides numerical stability, and $c_R,c_A>0$ are clipping thresholds. All valid tokens of a response share the same advantage $A_{x,g}$.

Using a PPO-style clipped objective~\cite{schulman2017ppo}, let $\rho=\pi_\theta(y_t\mid x,y_{<t})/\allowbreak\pi_{\mathrm{old}}(y_t\mid x,y_{<t})$ be the token probability ratio and $\bar\rho=\operatorname{clip}(\rho,1-\epsilon_{\mathrm{PPO}},1+\epsilon_{\mathrm{PPO}})$. The surrogate objective is $J=\min(\rho A,\bar\rho A)$. When $A<0$, dual clipping further applies $\max(J,c_{\mathrm{dual}}A)$, where $c_{\mathrm{dual}}>1$; otherwise, $J$ is retained. Denote the result by $J^{\mathrm{dual}}_{x,g,t}$.

For a minibatch of $B$ sampled responses, let $m_{x,g,t}$ be the response mask. The prefix-weighted policy loss is
\begin{equation}
 \mathcal L_{\mathrm{PG}}=-\frac1B\sum_{x,g}
 \frac{\sum_t m_{x,g,t}w_{x,g,t}J^{\mathrm{dual}}_{x,g,t}}
 {\sum_t m_{x,g,t}w_{x,g,t}}.
 \label{eq:weightedpg}
\end{equation}
Adding KL regularization relative to the fixed SFT reference policy gives the final training objective:
\begin{equation}
 \mathcal L_{\mathrm{actor}}=\mathcal L_{\mathrm{PG}}+\beta_{\mathrm{KL}}\mathcal L_{\mathrm{KL}}.
 \label{eq:actor}
\end{equation}
Here, $\beta_{\mathrm{KL}}>0$ is the regularization coefficient. $\mathcal L_{\mathrm{KL}}$ averages clipped KL-divergence estimates over valid tokens within each response and then across responses.

\begin{table*}[t]
\centering
\caption{Evaluation of Stage-3 SFT strategies. All variants are compared against the same SFT Base. (a) Absolute plain HR of the base and gains from the evaluated methods. (b) GMV and Q2I-GMV recall through Top-50 for the Q2I-example- and level-weighted checkpoint. All changes are in percentage points.}
\label{tab:sft}
\small
\begin{subtable}[t]{\textwidth}
\centering
\caption{Plain HR: absolute baseline (\%) and gains from the evaluated strategies (pp).}
\setlength{\tabcolsep}{3.4pt}
\begin{tabular*}{\linewidth}{@{\extracolsep{\fill}}llrrrrrr@{}}
\toprule
Direction & Method & @1 & @10 & @20 & @50 & @100 & @1000 \\
\midrule
Baseline & SFT Base (absolute, \%) & 6.39 & 26.33 & 34.68 & 46.19 & 54.59 & 72.84 \\
\midrule
Value & Q2I weight + level loss & -0.01 & \gain{+0.05} & \gain{+0.10} & \gain{+0.13} & \gain{+0.18} & \gain{+0.16} \\
Hierarchy & Level-weighted loss & -0.02 & 0.00 & +0.02 & +0.05 & +0.08 & \gain{+0.15} \\
Intent & Recent queries + global summary & \gain{+0.10} & \gain{+0.26} & \gain{+0.30} & \gain{+0.31} & \gain{+0.28} & \gain{+0.23} \\
Ordering & LO-SFT & +0.06 & +0.13 & +0.12 & +0.14 & +0.12 & +0.07 \\
\bottomrule
\end{tabular*}
\end{subtable}

\medskip
\begin{subtable}[t]{\textwidth}
\centering
\caption{Value-weighted recall (\%).}
\setlength{\tabcolsep}{4.0pt}
\begin{tabular*}{\linewidth}{@{\extracolsep{\fill}}rrrrrrr@{}}
\toprule
& \multicolumn{3}{c}{GMV Recall} & \multicolumn{3}{c}{Q2I-GMV Recall} \\
\cmidrule(lr){2-4}\cmidrule(lr){5-7}
$K$ & Base & Q2I + level & $\Delta$ & Base & Q2I + level & $\Delta$ \\
\midrule
1    & 30.88 & 33.06 & \gain{+2.18} & 36.12 & 42.35 & \gain{+6.23} \\
10   & 92.96 & 94.42 & \gain{+1.46} & 95.18 & 97.01 & \gain{+1.83} \\
20   & 97.57 & 97.73 & +0.16 & 98.06 & 99.06 & \gain{+1.00} \\
50   & 99.35 & 99.36 & +0.01 & 99.81 & 99.83 & +0.02 \\
\bottomrule
\end{tabular*}
\end{subtable}
\end{table*}

\begin{table}[b]
\centering
\caption{Q2I-GMV recall of SFT Base models. Absolute scores are percentages; gains are percentage points.}
\label{tab:sidvalue}
\small
\setlength{\tabcolsep}{4pt}
\begin{tabular*}{\linewidth}{@{\extracolsep{\fill}}lrrr@{}}
\toprule
Method & @1 & @10 & @20 \\
\midrule
RQ-VAE-$3$ & 20.62 & 81.53 & 97.13 \\
VARG-ID SFT Base & \textbf{36.12} & \textbf{95.18} & \textbf{98.06} \\
$\Delta$ (pp) & \gain{+15.50} & \gain{+13.65} & \gain{+0.93} \\
\bottomrule
\end{tabular*}
\end{table}

\begin{table*}[t]
\centering
\caption{Prefix-GRPO results versus its initialization model, SFT Base (\Cref{tab:sft}): head-level retrieval (left), value-weighted recall (right), and behavior-stratified HR@1 (bottom). Absolute scores are percentages; changes are percentage points.}
\label{tab:rl}
\small
\setlength{\tabcolsep}{5.3pt}
\begin{tabular}{lrrr@{\hspace{0.50in}}lrrr}
\toprule
Plain metric & SFT & Prefix-GRPO & $\Delta$ & Value metric & SFT & Prefix-GRPO & $\Delta$ \\
\midrule
HR@1 & 6.39 & 6.44 & +0.05 & GMV Recall@1 & 30.88 & 36.35 & \gain{+5.47} \\
HR@10 & 26.33 & 26.46 & +0.13 & GMV Recall@10 & 92.96 & 95.20 & \gain{+2.24} \\
HR@20 & 34.68 & 34.85 & \gain{+0.17} & GMV Recall@20 & 97.57 & 98.06 & +0.49 \\
HR@50 & 46.19 & 46.30 & \gain{+0.11} & Q2I-GMV Recall@1 & 36.12 & 38.60 & \gain{+2.48} \\
NDCG@10 & 15.15 & 15.30 & +0.15 & Q2I-GMV Recall@10 & 95.18 & 97.21 & \gain{+2.03} \\
NDCG@20 & 17.26 & 17.44 & \gain{+0.18} & Q2I-GMV Recall@20 & 98.06 & 99.39 & \gain{+1.33} \\
\midrule
\multicolumn{8}{c}{\textit{Behavior-stratified HR@1}} \\
Buy HR@1 & 13.51 & 13.87 & \gain{+0.36} & Click HR@1 & 5.97 & 6.00 & +0.03 \\
\bottomrule
\end{tabular}
\end{table*}

\begin{table}[t]
\centering
\caption{Progressive ablation of reward signals and prefix-aware token weighting. All scores are percentages; best scores are bold.}
\label{tab:rlablation}
\small
\setlength{\tabcolsep}{2pt}
\begin{tabular*}{\columnwidth}{@{\extracolsep{\fill}}lrrrrrr@{}}
\toprule
\multirow{2}{*}{Method} & \multicolumn{3}{c}{HR} & \multicolumn{2}{c}{Q2I-GMV Recall} & NDCG \\
\cmidrule(lr){2-4}\cmidrule(lr){5-6}\cmidrule(lr){7-7}
& @1 & @10 & @20 & @1 & @10 & @10 \\
\midrule
SFT Base & 6.39 & 26.33 & 34.68 & 36.12 & 95.18 & 15.15 \\
Behavior GRPO & 6.43 & 26.39 & 34.76 & 38.31 & 96.82 & 15.18 \\
Multi-source GRPO & 6.43 & 26.45 & 34.81 & 38.52 & 97.12 & 15.29 \\
Full Prefix-GRPO & \textbf{6.44} & \textbf{26.46} & \textbf{34.85} & \textbf{38.60} & \textbf{97.21} & \textbf{15.30} \\
\bottomrule
\end{tabular*}
\end{table}

\begin{table}[t]
\centering
\caption{Online A/B results over 14 days and 20\% of search traffic. All values are relative changes against the production control.}
\label{tab:online}
\small
\setlength{\tabcolsep}{2.2pt}
\begin{tabular*}{\columnwidth}{@{\extracolsep{\fill}}lrrrrrr@{}}
\toprule
Setting & GMV & IPV/user & Exp. PV/user & UCTR & UCTCVR & PCTR \\
\midrule
VARG & \gain{+1.45\%} & \gain{+0.22\%} & -0.09\% & \gain{+0.13\%} & \gain{+0.13\%} & \gain{+0.31\%}  \\
\bottomrule
\end{tabular*}
\end{table}

\subsection{Serving}
\label{sec:serving}
At inference, the model generates Top-$K$ SID paths using Trie-constrained beam search conditioned on the expanded context. The Trie is built from the published SID set. Let $\mathcal T(p)$ denote the set of legal child tokens after partial path $p$. Before each beam expansion, logits are masked as
\begin{equation}
 \widetilde z_v(p)=
 \begin{cases}
 z_v(p), & v\in\mathcal T(p),\\
 -\infty, & v\notin\mathcal T(p).
 \end{cases}
 \label{eq:triemask}
\end{equation}
At the third level, only occupied rank tokens satisfying $0\leq r<C_{ab}$ are allowed. Masking prunes invalid branches before beam expansion, so the generated complete SIDs can be mapped to valid items.

The Trie is rebuilt by the product-refresh loop and kept version-consistent with the SID-to-item map. Generated SIDs are expanded into item candidates through the version-matched map and directly admitted to the existing final ranker as an additional retrieval channel, augmenting the production multi-channel retrieval stack.


\section{Experiments}
\label{sec:experiments}

\subsection{Experimental Setup}
\paragraph{Dataset.}
All experiments were conducted on real search logs from Tmall App, a major mobile e-commerce platform. VARG-ID covers 51.43M products, with its tokenizer trained on query--item pairs and item-only examples. The three-stage generative model was trained on 51.43M samples for Item2SID, 48.00M for Query2SID, and 105.69M for personalized User--Query2SID. For Prefix-GRPO, we construct a training pool of approximately 1.03M records sampled from request-level data. Personalized evaluation and SID reordering audits used 3.85M samples; lightweight decoding was evaluated on a separate query-only set of 180,324 samples.

\paragraph{Metrics.}
HR@$K$ measures whether the target SID appears among the top-$K$ deduplicated predictions; NDCG@$K$ assesses ranking quality. GMV Recall and Q2I-GMV Recall measure the fraction of evaluation-sample value covered by these hits, weighted by the item's 30-day GMV and the query--item pair's 14-day GMV, respectively. The online A/B test evaluates a fixed candidate quota for direct admission over 14 days with 20\% of search traffic, reporting relative changes in business and engagement metrics against the production control.

\paragraph{Implementation Details.}
We extend Qwen2.5-0.5B-Instruct with SID tokens and perform full-parameter SFT on 128 GPUs. Learning rates are $10^{-4}$ for Stage~1, $4\times10^{-5}$ for Stages~2--3, and $5\times10^{-6}$ for LO-SFT. We evaluate SFT strategies with matched epoch budgets. Prefix-GRPO runs on 64 GPUs at a learning rate of $10^{-6}$, sampling eight responses per prompt with 1,024 prompts per rollout batch.

Hyperparameter settings are provided in Appendix~\ref{app:hyperparameters}.


\subsection{SID Observations: What Identifier Fits Recall-and-Pre-Ranking?}
\label{sec:sidobs}

\Cref{tab:sid}(a) shows that adding semantic quantization levels reduces collisions, but some items still share identifiers and decoding becomes longer. VARG-ID retains a two-level semantic prefix and distinguishes items within each prefix using a value-ordered third token, achieving collision-free addressing at initial construction. Both semantic codebooks are fully utilized, and the rank vocabulary has sufficient capacity for the largest observed item cluster.

\Cref{tab:sidvalue} shows that the VARG-ID SFT base improves Q2I-GMV Recall@1/10/20 by 15.50/13.65/0.93 percentage points over the SFT base using three-level RQ-VAE identifiers. These results indicate that VARG-ID achieves better value recall before RL, helping prioritize high-value items under limited candidate budgets.

With the SFT model fixed, full within-prefix reordering substantially reduces complete-SID HR while leaving semantic-prefix HR nearly unchanged, indicating that third-token label drift disrupts learned addressing relationships. These diagnostics support freezing addresses in the product loop. Detailed results and Trie-decoding diagnostics are provided in Appendix~\ref{app:sid-diagnostics}.


\subsection{SFT Strategies}
\label{sec:sftresults}

\paragraph{Value and hierarchy supervision.}
Joint Q2I and level weighting yields small gains in most reported HR metrics (\Cref{tab:sft}(a)). Value-recall gains are concentrated near the beam head: GMV Recall@1 and Q2I-GMV Recall@1 increase by 2.18 and 6.23 percentage points, respectively, while changes at Top-50 are small (\Cref{tab:sft}(b)). Level weighting also yields small HR gains at larger candidate budgets.

\paragraph{Intent expansion.}
Adding recent queries and the global-interest summary improves all reported HR metrics, with gains of 0.30 and 0.31 percentage points at HR@20 and HR@50, respectively. These results suggest that incorporating short-term intent and long-term preferences helps improve personalized retrieval.

\paragraph{Local ordinal supervision.}
LO-SFT yields small gains in all reported HR metrics, with HR@50 increasing from 46.19\% to 46.33\%. 

\subsection{Overall Prefix-GRPO Results}
\label{sec:rlresults}


Relative to the SFT model using VARG-ID, Prefix-GRPO improves head-level hit rate, ranking quality, and value recall (\Cref{tab:rl}). GMV Recall@1 and Q2I-GMV Recall@1 increase by 5.47 and 2.48 percentage points, respectively; HR@1 improves by 0.36 points for purchase-labeled requests and 0.03 points for click-labeled requests.

\paragraph{Reward and prefix ablation.}
\Cref{tab:rlablation} compares SFT Base with three RL configurations. All three retain the legality gates. Behavior GRPO uses user-behavior rewards for occupied SIDs; Multi-source GRPO adds ranker advantage and relevance rewards; Full Prefix-GRPO further enables prefix-aware token weighting. The first two configurations use unit weights for valid response tokens in the policy loss.

Relative to SFT Base, Behavior GRPO improves Q2I-GMV Recall@1/10 by 2.19/1.64 percentage points. Adding ranker and relevance feedback further improves these metrics by 0.21/0.30 points and NDCG@10 by 0.11 points. With the same reward, prefix weighting adds 0.08/0.09 points to Q2I-GMV Recall@1/10 and 0.04 points to HR@20. Behavioral feedback accounts for most of the observed head-value gain, ranker and relevance feedback further improve value recall and ranking, and prefix weighting provides small but consistent additional gains.

\subsection{Online A/B Test}
\label{sec:online}

\paragraph{Online system validation.}
We conducted a 14-day online A/B test in Tmall App covering 20\% of search traffic (\Cref{tab:online}). Both groups use the same total candidate budget for final ranking. \method uses Trie-constrained decoding with layer-wise beam widths of (20, 50, 500); generated item candidates bypass pre-ranking and enter the existing final ranker directly through a reserved quota. Compared with the production control, GMV increases by 1.45\%, per-user IPV by 0.22\%, UCTR and UCTCVR by 0.13\% each, and PCTR by 0.31\%. These results support the effectiveness of \method for integrated recall and pre-ranking.

The AI shopping-guide query evaluation is provided in Appendix~\ref{app:guide}.

\section{Conclusion}
\label{sec:conclusion}

\begingroup
We presented \method, a generative retrieval system for large-scale industrial e-commerce search that integrates recall and pre-ranking by directly admitting generated item candidates to the existing final ranker. VARG-ID uses bidirectional Q2I contrastive learning to align semantic prefixes with queries and stable EB-CVR ordering to assign the third token, providing collision-free, fine-grained item addresses and a business-value prior at initial construction. Three-stage SFT advances from item memorization to query-semantic and personalized retrieval, combining context expansion, value-aware and hierarchical supervision, and LO-SFT to improve candidate generation. Prefix-GRPO uses gated multi-source rewards and prefix-aware weighting to further align candidate priorities with business value and ranking objectives. Coordinated daily product and model updates preserve existing item addresses while incorporating new products and the latest behavioral feedback. Offline experiments evaluate identifier stability and demonstrate gains in retrieval quality and value recall. A 14-day online A/B test improves GMV by 1.45\% and PCTR by 0.31\%, supporting the effectiveness of the complete recall-and-pre-ranking system. Future work will explore quality--efficiency trade-offs under tighter candidate budgets and improve retrieval quality in lightweight decoding while retaining its efficiency gains.
\par\endgroup

\label{page:main-end}
\renewcommand{\bibfont}{\fontsize{7}{7.7}\selectfont}
\bibliographystyle{ACM-Reference-Format}
\bibliography{refs}


\begin{thebibliography}{35}


\ifx \showCODEN    \undefined \def \showCODEN     #1{\unskip}     \fi
\ifx \showISBNx    \undefined \def \showISBNx     #1{\unskip}     \fi
\ifx \showISBNxiii \undefined \def \showISBNxiii  #1{\unskip}     \fi
\ifx \showISSN     \undefined \def \showISSN      #1{\unskip}     \fi
\ifx \showLCCN     \undefined \def \showLCCN      #1{\unskip}     \fi
\ifx \shownote     \undefined \def \shownote      #1{#1}          \fi
\ifx \showarticletitle \undefined \def \showarticletitle #1{#1}   \fi
\ifx \showURL      \undefined \def \showURL       {\relax}        \fi
\providecommand\bibfield[2]{#2}
\providecommand\bibinfo[2]{#2}
\providecommand\natexlab[1]{#1}
\providecommand\showeprint[2][]{arXiv:#2}
\makeatletter
\@ifundefined{NAT@parse@date}{}{\let\NAT@parse@date@orig\NAT@parse@date}
\@ifundefined{NAT@parse@date}{}{\def\NAT@parse@date#1#2#3#4#5#6@@{\NAT@parse@date@orig#1#2#3#4#5#6@@\def\NAT@tempyear{0000}\def\NAT@tempexlab{{?}}\ifx\NAT@year\NAT@tempyear\ifx\NAT@exlab\NAT@tempexlab\def\NAT@date{[n.\,d.]}\else\edef\NAT@date{[n.\,d.]\NAT@exlab}\fi\fi}}
\makeatother

\bibitem[Bevilacqua et~al\mbox{.}(2022)]%
        {bevilacqua2022seal}
\bibfield{author}{\bibinfo{person}{Michele Bevilacqua},
  \bibinfo{person}{Giuseppe Ottaviano}, \bibinfo{person}{Patrick Lewis},
  \bibinfo{person}{Wen-tau Yih}, \bibinfo{person}{Sebastian Riedel}, {and}
  \bibinfo{person}{Fabio Petroni}.} \bibinfo{year}{2022}\natexlab{}.
\newblock \showarticletitle{Autoregressive Search Engines: Generating
  Substrings as Document Identifiers}.
\newblock \bibinfo{journal}{\emph{arXiv preprint arXiv:2204.10628}}
  (\bibinfo{year}{2022}).
\newblock
\urldef\tempurl%
\url{https://arxiv.org/abs/2204.10628}
\showURL{%
\tempurl}


\bibitem[Chen et~al\mbox{.}(2026)]%
        {chen2026onesearchv2}
\bibfield{author}{\bibinfo{person}{Ben Chen}, \bibinfo{person}{Siyuan Wang},
  \bibinfo{person}{Yufei Ma}, \bibinfo{person}{Zihan Liang},
  \bibinfo{person}{Xuxin Zhang}, \bibinfo{person}{Yue Lv},
  \bibinfo{person}{Ying Yang}, \bibinfo{person}{Huangyu Dai},
  \bibinfo{person}{Lingtao Mao}, \bibinfo{person}{Tong Zhao}, {et~al\mbox{.}}}
  \bibinfo{year}{2026}\natexlab{}.
\newblock \showarticletitle{{OneSearch-V2}: The Latent Reasoning Enhanced
  Self-distillation Generative Search Framework}.
\newblock \bibinfo{journal}{\emph{arXiv preprint arXiv:2603.24422}}
  (\bibinfo{year}{2026}).
\newblock


\bibitem[Chen et~al\mbox{.}(2023)]%
        {chen2023clever}
\bibfield{author}{\bibinfo{person}{Jiangui Chen}, \bibinfo{person}{Ruqing
  Zhang}, \bibinfo{person}{Jiafeng Guo}, \bibinfo{person}{Maarten de Rijke},
  \bibinfo{person}{Wei Chen}, \bibinfo{person}{Yixing Fan}, {and}
  \bibinfo{person}{Xueqi Cheng}.} \bibinfo{year}{2023}\natexlab{}.
\newblock \showarticletitle{Continual Learning for Generative Retrieval over
  Dynamic Corpora}. In \bibinfo{booktitle}{\emph{Proceedings of the ACM
  International Conference on Information and Knowledge Management}}.
\newblock
\href{https://doi.org/10.1145/3583780.3614821}{doi:\nolinkurl{10.1145/3583780.3614821}}


\bibitem[Cheng et~al\mbox{.}(2025)]%
        {cheng2025d2docid}
\bibfield{author}{\bibinfo{person}{Jiehan Cheng}, \bibinfo{person}{Zhicheng
  Dou}, \bibinfo{person}{Yutao Zhu}, {and} \bibinfo{person}{Xiaoxi Li}.}
  \bibinfo{year}{2025}\natexlab{}.
\newblock \showarticletitle{Descriptive and Discriminative Document Identifiers
  for Generative Retrieval}. In \bibinfo{booktitle}{\emph{Proceedings of the
  AAAI Conference on Artificial Intelligence}}, Vol.~\bibinfo{volume}{39}.
  \bibinfo{pages}{11518--11526}.
\newblock
\href{https://doi.org/10.1609/aaai.v39i11.33253}{doi:\nolinkurl{10.1609/aaai.v39i11.33253}}


\bibitem[De~Cao et~al\mbox{.}(2021)]%
        {decao2021genre}
\bibfield{author}{\bibinfo{person}{Nicola De~Cao}, \bibinfo{person}{Gautier
  Izacard}, \bibinfo{person}{Sebastian Riedel}, {and} \bibinfo{person}{Fabio
  Petroni}.} \bibinfo{year}{2021}\natexlab{}.
\newblock \showarticletitle{Autoregressive Entity Retrieval}. In
  \bibinfo{booktitle}{\emph{International Conference on Learning
  Representations}}.
\newblock
\urldef\tempurl%
\url{https://openreview.net/forum?id=5k8F6UU39V}
\showURL{%
\tempurl}


\bibitem[Deng et~al\mbox{.}(2025)]%
        {deng2025onerec}
\bibfield{author}{\bibinfo{person}{Jiaxin Deng}, \bibinfo{person}{Shiyao Wang},
  \bibinfo{person}{Kuo Cai}, \bibinfo{person}{Lejian Ren},
  \bibinfo{person}{Qigen Hu}, \bibinfo{person}{Weifeng Ding},
  \bibinfo{person}{Qiang Luo}, {and} \bibinfo{person}{Guorui Zhou}.}
  \bibinfo{year}{2025}\natexlab{}.
\newblock \showarticletitle{{OneRec}: Unifying Retrieve and Rank with
  Generative Recommender and Iterative Preference Alignment}.
\newblock \bibinfo{journal}{\emph{arXiv preprint arXiv:2502.18965}}
  (\bibinfo{year}{2025}).
\newblock
\urldef\tempurl%
\url{https://arxiv.org/abs/2502.18965}
\showURL{%
\tempurl}


\bibitem[Fu et~al\mbox{.}(2026)]%
        {fu2026forge}
\bibfield{author}{\bibinfo{person}{Kairui Fu}, \bibinfo{person}{Tao Zhang},
  \bibinfo{person}{Shuwen Xiao}, \bibinfo{person}{Ziyang Wang},
  \bibinfo{person}{Xinming Zhang}, \bibinfo{person}{Chenchi Zhang},
  \bibinfo{person}{Yuliang Yan}, \bibinfo{person}{Junjun Zheng},
  \bibinfo{person}{Xiangheng Kong}, \bibinfo{person}{Shengyu Zhang},
  \bibinfo{person}{Kun Kuang}, {and} \bibinfo{person}{Yuning Jiang}.}
  \bibinfo{year}{2026}\natexlab{}.
\newblock \showarticletitle{{FORGE}: Forming Semantic Identifiers for
  Generative Retrieval in Industrial Datasets}. In
  \bibinfo{booktitle}{\emph{Proceedings of the ACM SIGKDD Conference on
  Knowledge Discovery and Data Mining}}.
\newblock
\urldef\tempurl%
\url{https://arxiv.org/abs/2509.20904}
\showURL{%
\tempurl}


\bibitem[Geng et~al\mbox{.}(2022)]%
        {geng2022p5}
\bibfield{author}{\bibinfo{person}{Shijie Geng}, \bibinfo{person}{Shuchang
  Liu}, \bibinfo{person}{Zuohui Fu}, \bibinfo{person}{Yingqiang Ge}, {and}
  \bibinfo{person}{Yongfeng Zhang}.} \bibinfo{year}{2022}\natexlab{}.
\newblock \showarticletitle{Recommendation as Language Processing ({RLP}): A
  Unified Pretrain, Personalized Prompt \& Predict Paradigm ({P5})}. In
  \bibinfo{booktitle}{\emph{Proceedings of the ACM Conference on Recommender
  Systems}}.
\newblock
\urldef\tempurl%
\url{https://arxiv.org/abs/2203.13366}
\showURL{%
\tempurl}


\bibitem[Karpukhin et~al\mbox{.}(2020)]%
        {karpukhin2020dpr}
\bibfield{author}{\bibinfo{person}{Vladimir Karpukhin}, \bibinfo{person}{Barlas
  O{\u g}uz}, \bibinfo{person}{Sewon Min}, \bibinfo{person}{Patrick Lewis},
  \bibinfo{person}{Ledell Wu}, \bibinfo{person}{Sergey Edunov},
  \bibinfo{person}{Danqi Chen}, {and} \bibinfo{person}{Wen-tau Yih}.}
  \bibinfo{year}{2020}\natexlab{}.
\newblock \showarticletitle{Dense Passage Retrieval for Open-Domain Question
  Answering}.
\newblock \bibinfo{journal}{\emph{arXiv preprint arXiv:2004.04906}}
  (\bibinfo{year}{2020}).
\newblock
\urldef\tempurl%
\url{https://arxiv.org/abs/2004.04906}
\showURL{%
\tempurl}


\bibitem[Khattab and Zaharia(2020)]%
        {khattab2020colbert}
\bibfield{author}{\bibinfo{person}{Omar Khattab} {and} \bibinfo{person}{Matei
  Zaharia}.} \bibinfo{year}{2020}\natexlab{}.
\newblock \showarticletitle{{ColBERT}: Efficient and Effective Passage Search
  via Contextualized Late Interaction over {BERT}}. In
  \bibinfo{booktitle}{\emph{Proceedings of the International ACM SIGIR
  Conference on Research and Development in Information Retrieval}}.
\newblock
\urldef\tempurl%
\url{https://arxiv.org/abs/2004.12832}
\showURL{%
\tempurl}


\bibitem[Lee et~al\mbox{.}(2022)]%
        {lee2022rqvae}
\bibfield{author}{\bibinfo{person}{Doyup Lee}, \bibinfo{person}{Chiheon Kim},
  \bibinfo{person}{Saehoon Kim}, \bibinfo{person}{Minsu Cho}, {and}
  \bibinfo{person}{Wook-Shin Han}.} \bibinfo{year}{2022}\natexlab{}.
\newblock \showarticletitle{Autoregressive Image Generation Using Residual
  Quantization}. In \bibinfo{booktitle}{\emph{Proceedings of the IEEE/CVF
  Conference on Computer Vision and Pattern Recognition}}.
\newblock


\bibitem[Li et~al\mbox{.}(2024a)]%
        {li2024generativesurvey}
\bibfield{author}{\bibinfo{person}{Yongqi Li}, \bibinfo{person}{Xinyu Lin},
  \bibinfo{person}{Wenjie Wang}, \bibinfo{person}{Fuli Feng},
  \bibinfo{person}{Liang Pang}, \bibinfo{person}{Wenjie Li},
  \bibinfo{person}{Liqiang Nie}, \bibinfo{person}{Xiangnan He}, {and}
  \bibinfo{person}{Tat-Seng Chua}.} \bibinfo{year}{2024}\natexlab{a}.
\newblock \showarticletitle{A Survey of Generative Search and Recommendation in
  the Era of Large Language Models}.
\newblock \bibinfo{journal}{\emph{arXiv preprint arXiv:2404.16924}}
  (\bibinfo{year}{2024}).
\newblock
\urldef\tempurl%
\url{https://arxiv.org/abs/2404.16924}
\showURL{%
\tempurl}


\bibitem[Li et~al\mbox{.}(2024b)]%
        {li2024ltrgr}
\bibfield{author}{\bibinfo{person}{Yongqi Li}, \bibinfo{person}{Nan Yang},
  \bibinfo{person}{Liang Wang}, \bibinfo{person}{Furu Wei}, {and}
  \bibinfo{person}{Wenjie Li}.} \bibinfo{year}{2024}\natexlab{b}.
\newblock \showarticletitle{Learning to Rank in Generative Retrieval}. In
  \bibinfo{booktitle}{\emph{Proceedings of the AAAI Conference on Artificial
  Intelligence}}.
\newblock
\urldef\tempurl%
\url{https://arxiv.org/abs/2306.15222}
\showURL{%
\tempurl}


\bibitem[Liang et~al\mbox{.}(2025)]%
        {liang2025tbgrecall}
\bibfield{author}{\bibinfo{person}{Zida Liang}, \bibinfo{person}{Changfa Wu},
  \bibinfo{person}{Dunxian Huang}, \bibinfo{person}{Weiqiang Sun},
  \bibinfo{person}{Ziyang Wang}, \bibinfo{person}{Yuliang Yan},
  \bibinfo{person}{Jian Wu}, \bibinfo{person}{Yuning Jiang},
  \bibinfo{person}{Bo Zheng}, \bibinfo{person}{Ke Chen}, \bibinfo{person}{Silu
  Zhou}, {and} \bibinfo{person}{Yu Zhang}.} \bibinfo{year}{2025}\natexlab{}.
\newblock \showarticletitle{{TBGRecall}: A Generative Retrieval Model for
  E-commerce Recommendation Scenarios}.
\newblock \bibinfo{journal}{\emph{arXiv preprint arXiv:2508.11977}}
  (\bibinfo{year}{2025}).
\newblock
\urldef\tempurl%
\url{https://arxiv.org/abs/2508.11977}
\showURL{%
\tempurl}


\bibitem[Ling et~al\mbox{.}(2026)]%
        {ling2026crid}
\bibfield{author}{\bibinfo{person}{Gui Ling}, \bibinfo{person}{Zhihong Chen},
  \bibinfo{person}{Yu Li}, \bibinfo{person}{Tong Xiong},
  \bibinfo{person}{Kunhai Lin}, \bibinfo{person}{Kaixuan Zhang},
  \bibinfo{person}{Yuliang Yan}, \bibinfo{person}{Dan Ou},
  \bibinfo{person}{Haihong Tang}, {and} \bibinfo{person}{Bo Zheng}.}
  \bibinfo{year}{2026}\natexlab{}.
\newblock \showarticletitle{Beyond Semantic IDs: Encoding Business-Value
  Ranking into Document Identifiers for Generative Retrieval}.
\newblock \bibinfo{journal}{\emph{arXiv preprint arXiv:2607.11392}}
  (\bibinfo{year}{2026}).
\newblock


\bibitem[Malkov and Yashunin(2016)]%
        {malkov2016hnsw}
\bibfield{author}{\bibinfo{person}{Yu.~A. Malkov} {and} \bibinfo{person}{D.~A.
  Yashunin}.} \bibinfo{year}{2016}\natexlab{}.
\newblock \showarticletitle{Efficient and Robust Approximate Nearest Neighbor
  Search Using {Hierarchical Navigable Small World} Graphs}.
\newblock \bibinfo{journal}{\emph{arXiv preprint arXiv:1603.09320}}
  (\bibinfo{year}{2016}).
\newblock
\urldef\tempurl%
\url{https://arxiv.org/abs/1603.09320}
\showURL{%
\tempurl}


\bibitem[Mehta et~al\mbox{.}(2022)]%
        {mehta2022dsipp}
\bibfield{author}{\bibinfo{person}{Sanket~Vaibhav Mehta}, \bibinfo{person}{Jai
  Gupta}, \bibinfo{person}{Yi Tay}, \bibinfo{person}{Mostafa Dehghani},
  \bibinfo{person}{Vinh~Q. Tran}, \bibinfo{person}{Jinfeng Rao},
  \bibinfo{person}{Marc Najork}, \bibinfo{person}{Emma Strubell}, {and}
  \bibinfo{person}{Donald Metzler}.} \bibinfo{year}{2022}\natexlab{}.
\newblock \showarticletitle{{DSI++}: Updating Transformer Memory with New
  Documents}.
\newblock \bibinfo{journal}{\emph{arXiv preprint arXiv:2212.09744}}
  (\bibinfo{year}{2022}).
\newblock
\urldef\tempurl%
\url{https://arxiv.org/abs/2212.09744}
\showURL{%
\tempurl}


\bibitem[Rafailov et~al\mbox{.}(2023)]%
        {rafailov2023dpo}
\bibfield{author}{\bibinfo{person}{Rafael Rafailov}, \bibinfo{person}{Archit
  Sharma}, \bibinfo{person}{Eric Mitchell}, \bibinfo{person}{Stefano Ermon},
  \bibinfo{person}{Christopher~D. Manning}, {and} \bibinfo{person}{Chelsea
  Finn}.} \bibinfo{year}{2023}\natexlab{}.
\newblock \showarticletitle{Direct Preference Optimization: Your Language Model
  is Secretly a Reward Model}.
\newblock \bibinfo{journal}{\emph{arXiv preprint arXiv:2305.18290}}
  (\bibinfo{year}{2023}).
\newblock
\urldef\tempurl%
\url{https://arxiv.org/abs/2305.18290}
\showURL{%
\tempurl}


\bibitem[Rajput et~al\mbox{.}(2023)]%
        {rajput2023tiger}
\bibfield{author}{\bibinfo{person}{Shashank Rajput}, \bibinfo{person}{Nikhil
  Mehta}, \bibinfo{person}{Anima Singh}, \bibinfo{person}{Raghunandan~H.
  Keshavan}, \bibinfo{person}{Trung Vu}, \bibinfo{person}{Lukasz Heldt},
  \bibinfo{person}{Lichan Hong}, \bibinfo{person}{Yi Tay},
  \bibinfo{person}{Vinh~Q. Tran}, \bibinfo{person}{Jonah Samost},
  \bibinfo{person}{Maciej Kula}, \bibinfo{person}{Ed~H. Chi}, {and}
  \bibinfo{person}{Maheswaran Sathiamoorthy}.} \bibinfo{year}{2023}\natexlab{}.
\newblock \showarticletitle{Recommender Systems with Generative Retrieval}. In
  \bibinfo{booktitle}{\emph{Advances in Neural Information Processing
  Systems}}.
\newblock


\bibitem[Robertson and Zaragoza(2009)]%
        {robertson2009bm25}
\bibfield{author}{\bibinfo{person}{Stephen Robertson} {and}
  \bibinfo{person}{Hugo Zaragoza}.} \bibinfo{year}{2009}\natexlab{}.
\newblock \showarticletitle{The Probabilistic Relevance Framework: {BM25} and
  Beyond}.
\newblock \bibinfo{journal}{\emph{Foundations and Trends in Information
  Retrieval}} \bibinfo{volume}{3}, \bibinfo{number}{4} (\bibinfo{year}{2009}),
  \bibinfo{pages}{333--389}.
\newblock
\href{https://doi.org/10.1561/1500000019}{doi:\nolinkurl{10.1561/1500000019}}


\bibitem[Schulman et~al\mbox{.}(2017)]%
        {schulman2017ppo}
\bibfield{author}{\bibinfo{person}{John Schulman}, \bibinfo{person}{Filip
  Wolski}, \bibinfo{person}{Prafulla Dhariwal}, \bibinfo{person}{Alec Radford},
  {and} \bibinfo{person}{Oleg Klimov}.} \bibinfo{year}{2017}\natexlab{}.
\newblock \showarticletitle{Proximal Policy Optimization Algorithms}.
\newblock \bibinfo{journal}{\emph{arXiv preprint arXiv:1707.06347}}
  (\bibinfo{year}{2017}).
\newblock


\bibitem[Shao et~al\mbox{.}(2024)]%
        {shao2024deepseekmath}
\bibfield{author}{\bibinfo{person}{Zhihong Shao}, \bibinfo{person}{Peiyi Wang},
  \bibinfo{person}{Qihao Zhu}, \bibinfo{person}{Runxin Xu},
  \bibinfo{person}{Junxiao Song}, \bibinfo{person}{Xiao Bi},
  \bibinfo{person}{Haowei Zhang}, \bibinfo{person}{Mingchuan Zhang},
  \bibinfo{person}{Y.~K. Li}, \bibinfo{person}{Y. Wu}, {and}
  \bibinfo{person}{Daya Guo}.} \bibinfo{year}{2024}\natexlab{}.
\newblock \showarticletitle{{DeepSeekMath}: Pushing the Limits of Mathematical
  Reasoning in Open Language Models}.
\newblock \bibinfo{journal}{\emph{arXiv preprint arXiv:2402.03300}}
  (\bibinfo{year}{2024}).
\newblock


\bibitem[Sun et~al\mbox{.}(2023)]%
        {sun2023genret}
\bibfield{author}{\bibinfo{person}{Weiwei Sun}, \bibinfo{person}{Lingyong Yan},
  \bibinfo{person}{Zheng Chen}, \bibinfo{person}{Shuaiqiang Wang},
  \bibinfo{person}{Haichao Zhu}, \bibinfo{person}{Pengjie Ren},
  \bibinfo{person}{Zhumin Chen}, \bibinfo{person}{Dawei Yin},
  \bibinfo{person}{Maarten de Rijke}, {and} \bibinfo{person}{Zhaochun Ren}.}
  \bibinfo{year}{2023}\natexlab{}.
\newblock \showarticletitle{Learning to Tokenize for Generative Retrieval}. In
  \bibinfo{booktitle}{\emph{Advances in Neural Information Processing
  Systems}}.
\newblock
\urldef\tempurl%
\url{https://arxiv.org/abs/2304.04171}
\showURL{%
\tempurl}


\bibitem[Tan et~al\mbox{.}(2024)]%
        {tan2024idgenrec}
\bibfield{author}{\bibinfo{person}{Juntao Tan}, \bibinfo{person}{Shuyuan Xu},
  \bibinfo{person}{Wenyue Hua}, \bibinfo{person}{Yingqiang Ge},
  \bibinfo{person}{Zelong Li}, {and} \bibinfo{person}{Yongfeng Zhang}.}
  \bibinfo{year}{2024}\natexlab{}.
\newblock \showarticletitle{{IDGenRec}: {LLM-RecSys} Alignment with Textual
  {ID} Learning}. In \bibinfo{booktitle}{\emph{Proceedings of the International
  ACM SIGIR Conference on Research and Development in Information Retrieval}}.
\newblock
\urldef\tempurl%
\url{https://arxiv.org/abs/2403.19021}
\showURL{%
\tempurl}


\bibitem[Tay et~al\mbox{.}(2022)]%
        {tay2022dsi}
\bibfield{author}{\bibinfo{person}{Yi Tay}, \bibinfo{person}{Vinh~Q. Tran},
  \bibinfo{person}{Mostafa Dehghani}, \bibinfo{person}{Jianmo Ni},
  \bibinfo{person}{Dara Bahri}, \bibinfo{person}{Harsh Mehta},
  \bibinfo{person}{Zhen Qin}, \bibinfo{person}{Kai Hui}, \bibinfo{person}{Zhe
  Zhao}, \bibinfo{person}{Jai Gupta}, \bibinfo{person}{Tal Schuster},
  \bibinfo{person}{William~W. Cohen}, {and} \bibinfo{person}{Donald Metzler}.}
  \bibinfo{year}{2022}\natexlab{}.
\newblock \showarticletitle{Transformer Memory as a Differentiable Search
  Index}.
\newblock \bibinfo{journal}{\emph{arXiv preprint arXiv:2202.06991}}
  (\bibinfo{year}{2022}).
\newblock
\urldef\tempurl%
\url{https://arxiv.org/abs/2202.06991}
\showURL{%
\tempurl}


\bibitem[van~den Oord et~al\mbox{.}(2017)]%
        {oord2017vqvae}
\bibfield{author}{\bibinfo{person}{Aaron van~den Oord}, \bibinfo{person}{Oriol
  Vinyals}, {and} \bibinfo{person}{Koray Kavukcuoglu}.}
  \bibinfo{year}{2017}\natexlab{}.
\newblock \showarticletitle{Neural Discrete Representation Learning}. In
  \bibinfo{booktitle}{\emph{Advances in Neural Information Processing
  Systems}}.
\newblock
\urldef\tempurl%
\url{https://arxiv.org/abs/1711.00937}
\showURL{%
\tempurl}


\bibitem[Vaswani et~al\mbox{.}(2017)]%
        {vaswani2017attention}
\bibfield{author}{\bibinfo{person}{Ashish Vaswani}, \bibinfo{person}{Noam
  Shazeer}, \bibinfo{person}{Niki Parmar}, \bibinfo{person}{Jakob Uszkoreit},
  \bibinfo{person}{Llion Jones}, \bibinfo{person}{Aidan~N. Gomez},
  \bibinfo{person}{Lukasz Kaiser}, {and} \bibinfo{person}{Illia Polosukhin}.}
  \bibinfo{year}{2017}\natexlab{}.
\newblock \showarticletitle{Attention Is All You Need}. In
  \bibinfo{booktitle}{\emph{Advances in Neural Information Processing
  Systems}}.
\newblock


\bibitem[Wang et~al\mbox{.}(2024)]%
        {wang2024letter}
\bibfield{author}{\bibinfo{person}{Wenjie Wang}, \bibinfo{person}{Honghui Bao},
  \bibinfo{person}{Xinyu Lin}, \bibinfo{person}{Jizhi Zhang},
  \bibinfo{person}{Yongqi Li}, \bibinfo{person}{Fuli Feng},
  \bibinfo{person}{See-Kiong Ng}, {and} \bibinfo{person}{Tat-Seng Chua}.}
  \bibinfo{year}{2024}\natexlab{}.
\newblock \showarticletitle{Learnable Item Tokenization for Generative
  Recommendation}.
\newblock \bibinfo{journal}{\emph{arXiv preprint arXiv:2405.07314}}
  (\bibinfo{year}{2024}).
\newblock
\urldef\tempurl%
\url{https://arxiv.org/abs/2405.07314}
\showURL{%
\tempurl}


\bibitem[Wang et~al\mbox{.}(2022)]%
        {wang2022nci}
\bibfield{author}{\bibinfo{person}{Yujing Wang}, \bibinfo{person}{Yingyan Hou},
  \bibinfo{person}{Haonan Wang}, \bibinfo{person}{Ziming Miao},
  \bibinfo{person}{Shibin Wu}, \bibinfo{person}{Hao Sun}, \bibinfo{person}{Qi
  Chen}, \bibinfo{person}{Yuqing Xia}, \bibinfo{person}{Chengmin Chi},
  \bibinfo{person}{Guoshuai Zhao}, \bibinfo{person}{Zheng Liu},
  \bibinfo{person}{Xing Xie}, \bibinfo{person}{Hao~Allen Sun},
  \bibinfo{person}{Weiwei Deng}, \bibinfo{person}{Qi Zhang}, {and}
  \bibinfo{person}{Mao Yang}.} \bibinfo{year}{2022}\natexlab{}.
\newblock \showarticletitle{A Neural Corpus Indexer for Document Retrieval}. In
  \bibinfo{booktitle}{\emph{Advances in Neural Information Processing
  Systems}}.
\newblock
\urldef\tempurl%
\url{https://arxiv.org/abs/2206.02743}
\showURL{%
\tempurl}


\bibitem[Wu et~al\mbox{.}(2024)]%
        {wu2024multivector}
\bibfield{author}{\bibinfo{person}{Shiguang Wu}, \bibinfo{person}{Wenda Wei},
  \bibinfo{person}{Mengqi Zhang}, \bibinfo{person}{Zhumin Chen},
  \bibinfo{person}{Jun Ma}, \bibinfo{person}{Zhaochun Ren},
  \bibinfo{person}{Maarten de Rijke}, {and} \bibinfo{person}{Pengjie Ren}.}
  \bibinfo{year}{2024}\natexlab{}.
\newblock \showarticletitle{Generative Retrieval as Multi-Vector Dense
  Retrieval}. In \bibinfo{booktitle}{\emph{Proceedings of the International ACM
  SIGIR Conference on Research and Development in Information Retrieval}}.
\newblock
\urldef\tempurl%
\url{https://arxiv.org/abs/2404.00684}
\showURL{%
\tempurl}


\bibitem[Yang et~al\mbox{.}(2024)]%
        {qwen2025}
\bibfield{author}{\bibinfo{person}{An Yang}, \bibinfo{person}{Baosong Yang},
  \bibinfo{person}{Beichen Zhang}, \bibinfo{person}{Binyuan Hui},
  \bibinfo{person}{Bo Zheng}, \bibinfo{person}{Bowen Yu},
  \bibinfo{person}{Chengyuan Li}, \bibinfo{person}{Dayiheng Liu},
  \bibinfo{person}{Fei Huang}, \bibinfo{person}{Haoran Wei}, {et~al\mbox{.}}}
  \bibinfo{year}{2024}\natexlab{}.
\newblock \showarticletitle{{Qwen2.5} Technical Report}.
\newblock \bibinfo{journal}{\emph{arXiv preprint arXiv:2412.15115}}
  (\bibinfo{year}{2024}).
\newblock


\bibitem[Zheng et~al\mbox{.}(2024)]%
        {zheng2024lcrec}
\bibfield{author}{\bibinfo{person}{Bowen Zheng}, \bibinfo{person}{Yupeng Hou},
  \bibinfo{person}{Hongyu Lu}, \bibinfo{person}{Yu Chen},
  \bibinfo{person}{Wayne~Xin Zhao}, \bibinfo{person}{Ming Chen}, {and}
  \bibinfo{person}{Ji-Rong Wen}.} \bibinfo{year}{2024}\natexlab{}.
\newblock \showarticletitle{Adapting Large Language Models by Integrating
  Collaborative Semantics for Recommendation}. In
  \bibinfo{booktitle}{\emph{Proceedings of the IEEE International Conference on
  Data Engineering}}.
\newblock
\urldef\tempurl%
\url{https://arxiv.org/abs/2311.09049}
\showURL{%
\tempurl}


\bibitem[Zhou et~al\mbox{.}(2025)]%
        {zhou2025onerecv2}
\bibfield{author}{\bibinfo{person}{Guorui Zhou}, \bibinfo{person}{Hengrui Hu},
  \bibinfo{person}{Hongtao Cheng}, \bibinfo{person}{Huanjie Wang},
  \bibinfo{person}{Jiaxin Deng}, \bibinfo{person}{Jinghao Zhang},
  \bibinfo{person}{Kuo Cai}, \bibinfo{person}{Lejian Ren}, \bibinfo{person}{Lu
  Ren}, \bibinfo{person}{Liao Yu}, {et~al\mbox{.}}}
  \bibinfo{year}{2025}\natexlab{}.
\newblock \showarticletitle{{OneRec-V2} Technical Report}.
\newblock \bibinfo{journal}{\emph{arXiv preprint arXiv:2508.20900}}
  (\bibinfo{year}{2025}).
\newblock
\urldef\tempurl%
\url{https://arxiv.org/abs/2508.20900}
\showURL{%
\tempurl}


\bibitem[Zhou et~al\mbox{.}(2023)]%
        {zhou2023relevancefeedback}
\bibfield{author}{\bibinfo{person}{Yujia Zhou}, \bibinfo{person}{Zhicheng Dou},
  {and} \bibinfo{person}{Ji-Rong Wen}.} \bibinfo{year}{2023}\natexlab{}.
\newblock \showarticletitle{Enhancing Generative Retrieval with Reinforcement
  Learning from Relevance Feedback}. In \bibinfo{booktitle}{\emph{Proceedings
  of the 2023 Conference on Empirical Methods in Natural Language Processing}}.
  \bibinfo{publisher}{Association for Computational Linguistics},
  \bibinfo{pages}{12481--12490}.
\newblock
\href{https://doi.org/10.18653/v1/2023.emnlp-main.768}{doi:\nolinkurl{10.18653/v1/2023.emnlp-main.768}}


\bibitem[Zhu et~al\mbox{.}(2026)]%
        {zhu2026cq}
\bibfield{author}{\bibinfo{person}{Jianbo Zhu}, \bibinfo{person}{Xing Fang},
  \bibinfo{person}{Jing Wang}, \bibinfo{person}{Mingmin Jin},
  \bibinfo{person}{Bokang Wang}, \bibinfo{person}{Guangxin Song},
  \bibinfo{person}{Zhenyu Xie}, {and} \bibinfo{person}{Junjie Bai}.}
  \bibinfo{year}{2026}\natexlab{}.
\newblock \showarticletitle{Efficient Generative Retrieval for E-commerce
  Search with Semantic Cluster IDs and Expert-Guided RL}.
\newblock \bibinfo{journal}{\emph{arXiv preprint arXiv:2605.14434}}
  (\bibinfo{year}{2026}).
\newblock


\end{thebibliography}

\clearpage
\noindent\textbf{Roadmap.} The appendix is organized as follows:
\begin{itemize}
  \item \hyperref[app:hyperparameters]{\textcolor{varggreen}{Appendix~\ref*{app:hyperparameters}}} provides hyperparameter settings for VARG-ID, supervised fine-tuning, and Prefix-GRPO.
  \item \hyperref[app:sid-diagnostics]{\textcolor{varggreen}{Appendix~\ref*{app:sid-diagnostics}}} reports SID stability and Trie-decoding diagnostics, additional LO-SFT analysis, and shopping-guide query evaluations.
  \item \hyperref[sec:lightdecode]{\textcolor{varggreen}{Appendix~\ref*{sec:lightdecode}}} describes the exploratory lightweight decoder and its efficiency--effectiveness trade-off.
  \item \hyperref[sec:case]{\textcolor{varggreen}{Appendix~\ref*{sec:case}}} presents personalized and cold-start retrieval case studies.
\end{itemize}

\appendix

\predisplaypenalty=0
\begin{table*}[t]
\centering
\caption{Evaluation of SIDs and constrained decoding on production data. (a) Initial identifier capacity and collision. (b) Full-reordering stability. (c) Trie-constrained decoding effectiveness and overhead.}
\label{tab:sid}
\small
\begin{subtable}[t]{0.56\textwidth}
\centering
\caption{Initial identifier capacity and collision.}
\setlength{\tabcolsep}{3pt}
\begin{tabular*}{\linewidth}{@{\extracolsep{\fill}}lrrrrrr@{}}
\toprule
Identifier & Tokens & Unique & Exclusive & Collision & Avg. & P99 / Max \\
\midrule
RQ-VAE-$3$ & 3 & 16.73\% & 8.75\% & 91.25\% & 5.98 & 76 / 7,580 \\
RQ-VAE-$4$ & 4 & 51.84\% & 38.42\% & 61.58\% & 1.93 & 15 / 2,364 \\
RQ+OPQ & 5 & 81.21\% & 71.62\% & 28.38\% & 1.23 & 5 / 968 \\
VARG prefix & 2 & 10.74\% & 4.32\% & 95.68\% & 9.31 & 129 / 4,221 \\
\textbf{VARG-ID} & 3 & \textbf{100\%} & \textbf{100\%} & \textbf{0\%} & \textbf{1.00} & \textbf{1 / 1} \\
\bottomrule
\end{tabular*}
\end{subtable}\hfill
\begin{subtable}[t]{0.42\textwidth}
\centering
\caption{Full-reordering stability.}
\setlength{\tabcolsep}{2.5pt}
\begin{tabular*}{\linewidth}{@{\extracolsep{\fill}}lrrrrr@{}}
\toprule
Policy & Change & Avg. shift & P90 & P99 & Top20 keep \\
\midrule
\textbf{EB-CVR} & \textbf{69.54\%} & \textbf{22.00} & \textbf{51} & \textbf{291} & \textbf{93.31\%} \\
Click & 77.61\% & 26.58 & 61 & 395 & 91.29\% \\
Volume & 76.93\% & 26.99 & 59 & 389 & 91.97\% \\
\midrule
Frozen update & 0.00\% & 0.00 & 0 & 0 & 100.00\% \\
\bottomrule
\end{tabular*}
\end{subtable}

\medskip
\begin{subtable}[t]{\textwidth}
\centering
\caption{Trie-constrained decoding.}
\setlength{\tabcolsep}{4pt}
\begin{tabular*}{\linewidth}{@{\extracolsep{\fill}}lrrrrrr@{}}
\toprule
Method & Legal rate (\%) & $\Delta$HR@20 & $\Delta$HR@100 & $\Delta$HR@1000 & Inference time $\Delta$ & Memory $\Delta$ \\
\midrule
Trie & $93.73\to\mathbf{100}$ & +0.0915 & \gain{+0.2343} & \gain{+0.3857} & +3.16\% & +0.13 GiB \\
\bottomrule
\end{tabular*}
\end{subtable}
\par\smallskip
{\footnotesize\raggedright
\textit{Notes.} (a) For $N$ items and $U$ occupied SIDs, Unique=$U/N$ and Avg.=$N/U$. Exclusive is the fraction of items with an unshared SID, and Collision=$1-\mathrm{Exclusive}$; P99/Max are the 99th percentile and maximum item counts per occupied SID.
(b) Change is the fraction of items whose third token changes; Avg.\ shift is mean absolute rank displacement, and P90/P99 are displacement quantiles. Top20 keep is the retention rate of old Top-20 items in the same prefix's new Top-20.
(c) HR gains are in percentage points. Time is cumulative forward-pass and beam-search time, and memory is peak GPU memory usage; both are measured on rank~0.\par}
\end{table*}

\begin{table}[!t]
\centering
\caption{Relevance Good rate (\%) in two independent evaluations of AI shopping-guide queries under different candidate-quota configurations.}
\label{tab:guide}
\small
\setlength{\tabcolsep}{3pt}
\begin{tabular*}{\columnwidth}{@{\extracolsep{\fill}}lrr@{}}
\toprule
System & Eval. 1 & Eval. 2 \\
\midrule
Orig. pipeline (largest quota) & 76.71 & \textbf{76.81} \\
Orig. pipeline (reduced quota) & 75.77 & 75.73 \\
Gen. baseline (DA) & 75.57 & -- \\
Gen. baseline (DA, larger quota) & 75.90 & 76.03 \\
\textbf{VARG (DA, smaller quota)} & \gain{76.77} & \gain{76.58} \\
\bottomrule
\end{tabular*}
\par\smallskip
{\footnotesize\raggedright Eval.\ 1/2: independent evaluation rounds. DA: direct admission to the final ranker, bypassing pre-ranking.\par}
\end{table}

\begin{figure}[b]
\centering
\includegraphics[width=0.95\columnwidth]{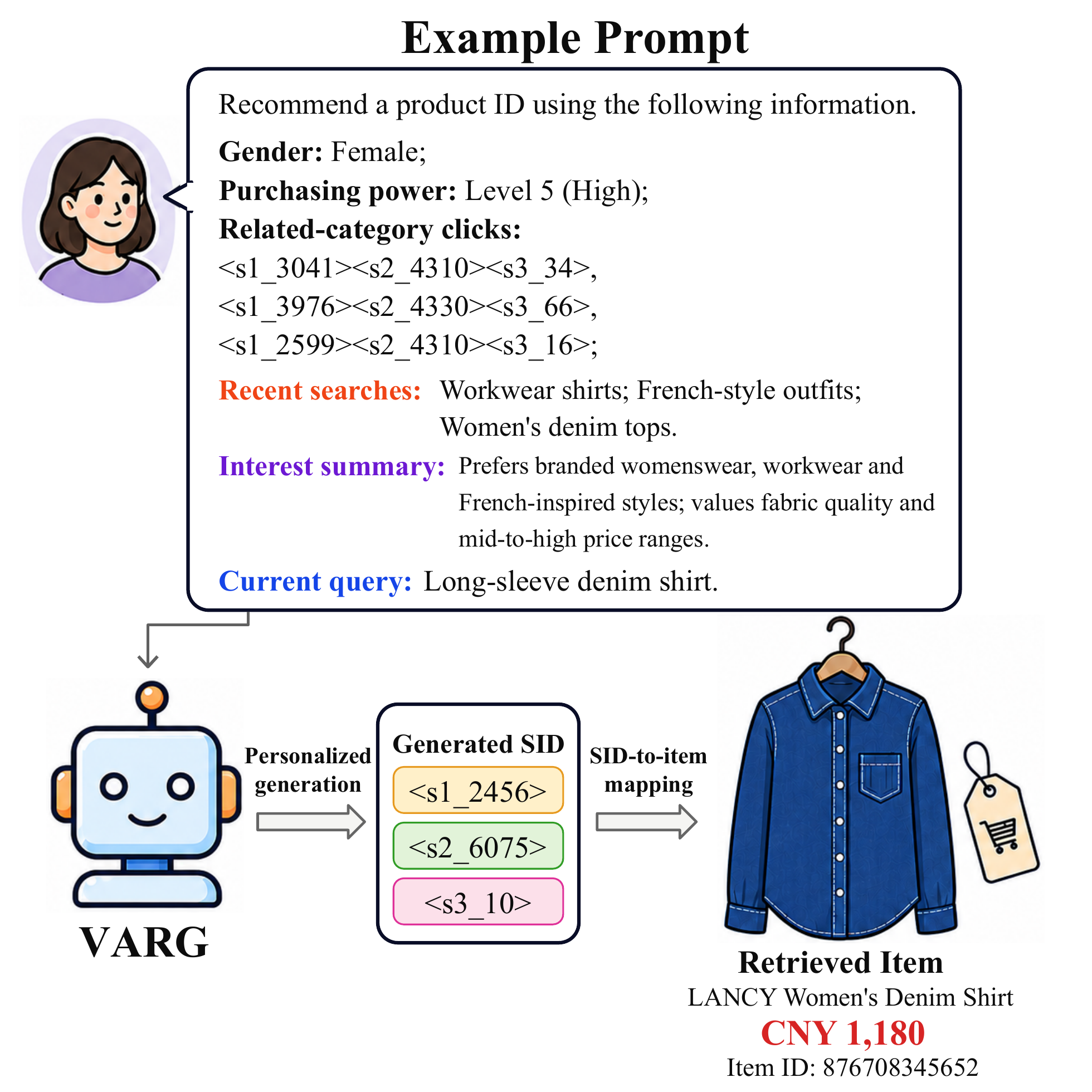}
\caption{Schematic personalized prompt-to-item example (English translation).}
\Description{Schematic personalized prompt-to-item example (English translation).}
\label{fig:promptexample}
\end{figure}

\begin{figure*}[!t]
\centering
\includegraphics[width=0.85\textwidth]{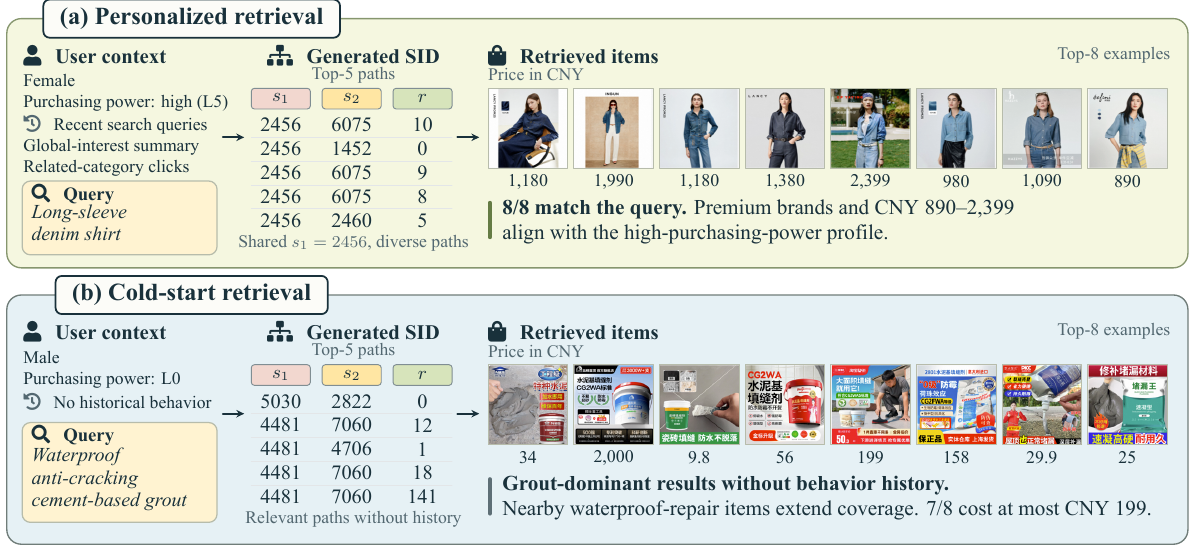}
\caption{Personalized and cold-start retrieval examples. Case A shows personalized retrieval with user context; Case B shows cold-start retrieval without behavioral history. Prices are in CNY.}
\Description{Personalized and cold-start retrieval examples. Case A shows personalized retrieval with user context; Case B shows cold-start retrieval without behavioral history. Prices are in CNY.}
\label{fig:case}
\end{figure*}

\section{Hyperparameter Settings}
\label{app:hyperparameters}
For VARG-ID, both semantic codebooks and the third-level rank vocabulary have 8,192 entries ($K_1=K_2=K_3=8192$). Reconstruction and commitment losses each have unit weight; the bidirectional Q2I contrastive loss uses $\lambda_{\mathrm{QI}}=0.001$ and temperature $\tau_{\mathrm{QI}}=0.1$, with at most 128 valid query--item pairs sampled per local minibatch.

For SFT strategies, the Q2I prior strength is $\alpha_q=5$, and intent expansion includes the three most recent search queries ($L_{\mathrm{recent}}=3$). LO-SFT uses neighborhood radius $R_{\mathrm{LO}}=3$, distance-decay temperature $\tau_{\mathrm{LO}}=1.0$, and auxiliary loss coefficient $\lambda_{\mathrm{LO}}=0.02$.

For Prefix-GRPO, malformed and unoccupied outputs receive rewards of $-2.0$ and $-1.0$, respectively. Purchase, click, and exposure rewards are $3.0$, $1.0$, and $0.1$. The ranker-advantage reward uses $\beta_{\mathrm{adv}}=0.2$ and $\tau_{\mathrm{adv}}=1.5$; relevance grades 3, 2, and 1 contribute $+0.08$, $+0.03$, and $-0.10$ for occupied SIDs without matched user actions. The prefix-weight floor is $d=0.1$, also used for EOS and control tokens, and the KL regularization coefficient is $\beta_{\mathrm{KL}}=0.8$.

\section{Supplementary Experiments}
\label{app:sid-diagnostics}
\label{app:guide}
EB-CVR is the most stable of the value-based reordering policies in \Cref{tab:sid}(b): compared with Click, it yields fewer third-token changes, smaller rank displacements, and higher Top-20 retention. We further evaluate how identifier changes affect the trained model's retrieval performance. We keep the SFT model parameters fixed and reorder all items within their semantic prefixes. After reordering, L1 HR@100 remains nearly constant (92.08\%$\to$92.00\%) and L1+L2 HR remains stable (71.80\%$\to$71.56\%), but complete-SID HR@100 drops by \loss{14.45 points}, and third-token teacher-forcing accuracy falls from 36.75\% to 21.24\%. This degradation reflects temporal drift in SID labels: as item statistics change, within-prefix reordering changes the third token assigned to the same item, creating a mismatch with the addressing relationships learned by the model. We therefore adopt frozen incremental updates to preserve existing item addresses.

During daily product updates, existing item addresses remain unchanged, while incremental insertion permits controlled SID sharing. This update policy incorporates new items while preserving the addressing relationships learned by the model.

\Cref{tab:sid}(c) shows that Trie constraints improve SID legality and retrieval hit rates with only small increases in inference time and memory usage.

\paragraph{Additional LO-SFT diagnostics.}
\label{app:lo-diagnostic}
Further analysis shows that NDCG@100 increases from 20.91\% to 21.00\%, while third-token top-1 accuracy, conditioned on the gold two-level semantic prefix, increases from 36.74\% to 37.02\%.

\paragraph{AI shopping-guide query evaluation.}
We conducted two independent relevance evaluations on 6K+ AI shopping-guide queries using the online search system. Good rate is the fraction of evaluated query--item pairs judged relevant. \Cref{tab:guide} shows that the original pipeline's Good rate decreases when its candidate quota is reduced. With a smaller candidate quota, \method outperforms the generative direct-admission baselines and achieves a Good rate comparable to or slightly higher than that of the original pipeline at its largest quota. These results indicate that \method maintains competitive relevance under a smaller candidate budget.

\section{Exploratory Lightweight Decoder}
\label{sec:lightdecode}
\paragraph{Shared representation and prediction heads.}
The lightweight decoder produces a complete SID with a single Transformer backbone forward pass, without autoregressive backbone decoding over successive SID tokens. We compute the prompt-end hidden state $h$ once and reuse it across levels and candidate paths. The remaining decisions are handled by lightweight, prefix-conditioned classification heads:
\begin{align}
 p_1(s_1\mid h)&=\operatorname{softmax}\!\left(g_1(h)\right),\notag\\
 p_2(s_2\mid h,s_1)&=\operatorname{softmax}\!\left(g_2([h;e_1(s_1)])\right),\notag\\
 p_3(s_3\mid h,s_1,s_2)&=\operatorname{softmax}\!\left(g_3([h;e_1(s_1);e_2(s_2)])\right),
 \label{eq:lightheads}
\end{align}
Here $s_3$ denotes the rank token $r$ used in the main text, $e_l$ is the embedding of the SID token at level $l$, and $g_l$ is the corresponding prediction head. The semicolon in $[h;e_1(s_1)]$ denotes vector concatenation. The first head predicts $s_1$ from $h$; the second augments $h$ with the selected $s_1$ embedding; and the third additionally incorporates the $s_2$ embedding to predict $r$. Thus, SID-level conditional dependencies are retained within the prediction heads, while the shared backbone representation remains fixed throughout decoding.

\paragraph{Teacher forcing and supervision.}
During training, the second head receives the ground-truth $s_1$, and the third receives the ground-truth pair $(s_1,s_2)$. Each head is supervised by cross-entropy against its corresponding SID token. All three heads share the prompt representation, so predicted SID tokens need not be fed back through the Transformer to construct later-level inputs. For $L=3$ levels, we normalize the loss coefficients $\alpha_l>0$ to obtain the training objective
\begin{equation}
 \widetilde\alpha_l=\frac{L\alpha_l}{\sum_{j=1}^{L}\alpha_j},\qquad
 \mathcal L_{\mathrm{head}}=\sum_{l=1}^{L}\widetilde\alpha_l
 \operatorname{CE}\!\left(p_l,s_l\right).
 \label{eq:lightloss}
\end{equation}
The normalized coefficients satisfy $\sum_{l=1}^{3}\widetilde\alpha_l=3$. They control the relative contribution of the three classification losses while fixing the sum of their coefficients; equal coefficients recover an equally weighted sum of the per-level losses.

\paragraph{Hierarchical beam search and computation.}
At inference, hierarchical beam expansion proceeds across the prediction heads, with prefix masks retaining occupied VARG-ID child tokens. We first compute and cache $h$, and retain high-scoring first-level tokens. For each retained path, the selected prefix embeddings are concatenated with the same $h$ to compute the next-level distribution. Path scores accumulate the conditional log probabilities across levels, and beam pruning retains the highest-scoring partial paths before the next expansion. After the third level, the search returns the Top-$K$ complete SIDs.

Candidate paths at the same level can be evaluated in a batch, while decisions across levels remain conditioned on the selected prefixes. Only one backbone forward pass is needed for the prompt; subsequent expansion and scoring use the lightweight heads. The computation still includes head evaluation and beam search, but does not require additional autoregressive Transformer passes for the SID tokens. \Cref{sec:efficiency} evaluates the decoder's efficiency--retrieval-quality trade-off; it remains an exploratory approach and is not used in the main VARG serving pipeline.

\balance
\subsection{Efficiency--Effectiveness Trade-off}
\label{sec:efficiency}

Retrieval quality was evaluated on the separate query-only set of 180,324 samples. We benchmarked the standard autoregressive decoder and the lightweight decoder in \Cref{sec:lightdecode} using PyTorch and Transformers, without an additional inference engine. Our pilot study shows that, relative to standard autoregressive decoding, the lightweight decoder reduces mean latency from 10.41 to 0.94 ms and achieves approximately $11.1\times$ the single-card QPS, but HR@1/20/100 drops by 1.13/8.38/10.77 percentage points. The lightweight decoder is not yet used in the main online \method pipeline and requires further improvement in retrieval quality before deployment.

\section{Case Study}
\label{sec:case}

Figure~\ref{fig:case} presents personalized retrieval (Case A) and cold-start retrieval (Case B). In Case A, recent queries reflect short-term intent, the global-interest summary captures long-term preferences, and related-category clicks provide behavioral evidence. All eight displayed items match the query ``long-sleeve denim shirt''; the premium brands and price range are consistent with the user's high purchasing power, illustrating personalized retrieval aligned with user preferences. Case B contains no behavioral history. For the query ``waterproof anti-cracking cement-based grout,'' the displayed results mainly comprise grout products and related waterproof-repair items, showing that retrieval remains semantically aligned with the query in this cold-start example.

\end{document}